\documentclass[aip, reprint, amsmath, amssymb]{revtex4-1}

\usepackage{amsmath}
\usepackage{amssymb}
\usepackage{amsfonts}
\usepackage{amsthm}
\usepackage{amscd}
\usepackage[T1]{fontenc}
\usepackage{mathrsfs}
\usepackage{latexsym}
\usepackage{physics}
\usepackage{graphicx}
\usepackage{dcolumn}
\usepackage{bm}
\usepackage[position=top, labelfont=bf]{subfig}

\usepackage[utf8]{inputenc}
\usepackage[T1]{fontenc}
\usepackage{mathptmx}
\DeclareMathAlphabet{\mathcal}{OMS}{cmsy}{m}{n} 

\usepackage{ragged2e}
\usepackage{float}
\usepackage{caption}
\usepackage{hyperref}

\makeatletter
\renewcommand\@make@capt@title[2]{%
	\@ifx@empty\float@link{\@firstofone}{\expandafter\href\expandafter{\float@link}}%
	{\textbf{#1}}\@caption@fignum@sep#2\quad}%
\makeatother

\makeatletter 
\renewcommand{\fnum@figure}{\textbf{Figure~\thefigure}}
\makeatother

\usepackage{bbm}

\draft 

\begin{document}

\title{Deep Learning for Automated Classification and Characterization of Amorphous Materials}

\author{Kirk Swanson} 
 \email{swansonk1@uchicago.edu}
 \affiliation{Pritzker School for Molecular Engineering, The University of Chicago, Chicago, IL 60637}
 \affiliation{Department of Computer Science, The University of Chicago, Chicago, IL 60637}
\author{Shubhendu Trivedi}
 \email{shubhendu@csail.mit.edu}
 \affiliation{Computer Science and Artificial Intelligence Laboratory, MIT, Cambridge, MA, 02139}
 \affiliation{Institute for Computational and Experimental Research in Mathematics, Brown University, Providence, RI 02903}
\author{Joshua Lequieu}
 \email{lequieu@mrl.ucsb.edu}
 \affiliation{Pritzker School for Molecular Engineering, The University of Chicago, Chicago, IL 60637}
 \affiliation{Materials Research Laboratory, University of California, Santa Barbara, CA 93106}
\author{Kyle Swanson}
 \email{swansonk@mit.edu}
 \affiliation{Computer Science and Artificial Intelligence Laboratory, MIT, Cambridge, MA, 02139}
 \author{Risi Kondor}
 \email{risi@cs.uchicago.edu}
 \affiliation{Department of Computer Science, The University of Chicago, Chicago, IL 60637}
  \affiliation{Department of Statistics, The University of Chicago, Chicago, IL 60637}
   \affiliation{Center for Computational Mathematics, Flatiron Institute, New York, NY 10010}

\date{\today}

\begin{abstract}
It is difficult to quantify structure-property relationships and to identify structural features of complex materials. The characterization of amorphous materials is especially challenging because their lack of long-range order makes it difficult to define structural metrics. In this work, we apply deep learning algorithms to accurately classify amorphous materials and characterize their structural features. Specifically, we show that convolutional neural networks and message passing neural networks can classify two-dimensional liquids and liquid-cooled glasses from molecular dynamics simulations with greater than 0.98 AUC, with no \textit{a priori} assumptions about local particle relationships, even when the liquids and glasses are prepared at the same inherent structure energy. Furthermore, we demonstrate that message passing neural networks surpass convolutional neural networks in this context in both accuracy and interpretability. We extract a clear interpretation of how message passing neural networks evaluate liquid and glass structures by using a self-attention mechanism. Using this interpretation, we derive three novel structural metrics that accurately characterize glass formation. The methods presented here provide us with a procedure to identify important structural features in materials that could be missed by standard techniques and give us a unique insight into how these neural networks process data.

\end{abstract}


\maketitle 

\section{Introduction}\label{sec:sec1}
Classifying material structures and predicting their properties are important tasks in materials science. The behavior of materials often depends strongly on their underlying structure, and understanding these structure-property relationships relies on accurately describing the structural features of a material. However, quantifying structure-property relationships and identifying structural features in complex materials are difficult tasks.  
\par A variety of standard techniques have been developed to analyze material structures. Some of the most common techniques include the Steinhardt bond order parameters, \cite{ST} Bond Angle Analysis (BAA), \cite{AJ} and Common Neighbor Analysis (CNA), \cite{HA} which are useful for detecting order-disorder transitions and differentiating between crystal structures in ordered samples.\cite{FG, D} Radial distribution functions, which measure a spatial average of interparticle distances, are also widely used for analyzing different materials and phases.\cite{Tuckerman} 
\par However, these standard techniques have limitations, especially for analyzing weakly crystalline or amorphous materials. As discussed in Reinhardt \textit{et al.},\cite{FG} the Steinhardt bond order parameters can be stymied by thermal fluctuations or ambiguous crystal structures. BAA relies on a small set of crystalline reference structures that may not be present in amorphous samples. CNA is more flexible than BAA, but it cannot provide accurate information about particles that do not exhibit known symmetries, making analysis of irregular structures challenging. Radial distribution functions rely on spatial averaging, which interferes with the ability to characterize complex anisotropic structures. 
\par To overcome some of these limitations, recent studies have focused on developing machine learning algorithms to automate material structure characterization. Many of these studies have concentrated on using supervised machine learning for crystal structure identification and have shown improvements over standard techniques. Geiger and Dellago and Dietz \textit{et al.},\cite{GD, D} for example, classify crystalline structures in polymorphic and mixed phase systems more accurately by using a set of local structure functions that are fed into a neural network. Ziletti \textit{et al.}\cite{Z} use convolutional neural networks to automate the process of classifying crystal symmetries, even in the presence of substantial defects, using images of computationally-generated diffraction patterns. Similar supervised learning methods have been used to analyze a variety of other ordered systems, including graphene. \cite{M, RK, LP, P}
\par Other studies have concentrated on crystal structure identification using unsupervised learning techniques, such as dimensionality-reduction algorithms. These unsupervised methods allow for the discovery of previously unidentified local structural features, a task that is often out of reach for standard methods and many supervised learning algorithms. For example, Reinhardt \textit{et al.}\cite{FG} use diffusion maps and clustering to identify new local structures in colloidal crystals, and Spellings and Glotzer\cite{SG} use spherical harmonic functions in conjunction with Gaussian mixture models to automatically identify different crystalline arrangements without relying on a set of reference structures. Unsupervised learning has also been used to study systems ranging from Lennard-Jones crystals to proteins. \cite{GC, ZM, LN, HB, YL, MG}
\par Although these machine learning approaches have shown substantial improvements over standard techniques when applied to a variety of systems, many of them still rely on the development of complex, hand-crafted descriptors of local particle environments. The results of machine learning analyses can be quite sensitive to the definitions of these descriptors. \cite{GC} Moreover, most of these studies have focused on characterization of crystalline or semi-crystalline structures, with little attention given to analysis of completely amorphous systems such as glassy materials. 
\par The characterization of glassy systems is especially challenging because their lack of long-range order makes it difficult to define  structural metrics. Nevertheless, several recent studies have been able to manually identify structural metrics for glassy systems. For example, Hu \textit{et al.}\cite{N} defined a metric, the average degree of local five-fold symmetry, that can differentiate between configurations of metallic liquids and glasses and that has a quantitative relationship with dynamics during glass formation Reid \textit{et al.}\cite{R} use a similar five-fold symmetry metric, based on the spherical harmonic functions, to compare the structures of two-dimensional liquid-cooled and vapor-deposited glasses.
\par There have also been some recent studies that have successfully used machine learning to uncover previously unknown relationships between structure and dynamics in glassy materials. \cite{DS, EC1, EC2, MX, EC3, SS, SS2, I, MH} In these approaches, a supervised machine learning algorithm, called the support vector machine, is used to define a metric, called "softness," that identifies populations of particles that are likely to dynamically rearrange. In this context, "softness" is used to link structure and dynamics, but it is not used to directly identify local structural features or to classify different material structures. 
\par In this work, we use deep learning to accurately classify amorphous materials and to derive new metrics that characterize their structures. We use two-dimensional liquids and liquid-cooled glasses generated by molecular dynamics simulations as archetypal examples of amorphous materials. Our classification algorithms make no \textit{a priori} assumptions about local relationships between particles and are not dependent on complex hand-crafted descriptors that define local particle environments. 
\par We explore the application of two different types of deep learning algorithms: convolutional neural networks and message passing neural networks. Convolutional neural networks have been used in a variety of material classification tasks, \cite{Z, M, RK, P} including the classification of ordered and disordered configurations from simulations of the Ising model, \cite{IM1, IM2, IM3} but to the best of our knowledge they have never before been used to classify different amorphous material structures from molecular dynamics simulations. By rendering particle configurations of liquids and glasses as two-dimensional images, we are able to distinguish between them with high accuracy using convolutional neural networks. However, there are several limitations that accompany the use of convolutional neural networks, including the potential introduction of artifacts via image rendering and limited interpretability. 
\par We overcome these issues by using message passing neural networks, which operate directly on the Cartesian coordinates of particles by representing a configuration of particles as a graph. Message passing neural networks have been used previously to predict properties of molecules, such as toxicity and solubility, \cite{S, GG, PJ, STR, STR2, SCH, PBJ, KKM} and to predict properties of crystals, such as formation energy and shear modulus. \cite{X} However, to the best of our knowledge, message passing neural networks have never before been used to classify different amorphous material structures from molecular dynamics simulations. By representing particle configurations of liquids and glasses as graphs, we are able to distinguish between them with high accuracy. Moreover, by using a technique called self-attention, we are able to extract an interpretation of how message passing neural networks evaluate liquid and glass configurations. Using this interpretation, we derive three novel structural metrics that characterize glass formation and that can differentiate between liquid and glass configurations without the use of machine learning. This not only provides us with a general method for identifying important structural features in amorphous materials, but it also gives us a unique insight into how these neural networks process data. This result provides clear proof of concept that message passing neural networks could be used in more complex and demanding classification and characterization tasks that stymie standard techniques. 
\section{Methods}\label{sec:sec2}
\subsection{Simulation Details}\label{sec:sec21}
\par Glasses are kinetically arrested states of matter which are generally prepared by cooling a liquid to temperatures below the glass transition, $T_g$, of the corresponding bulk material. \cite{R} When cooling is sufficiently rapid, the system avoids crystallization and instead solidifies into a glass, an amorphous state which has an atomic structure similar to that of a liquid but with the mechanical properties of a solid. The specific properties of these liquid-cooled glasses depend on the rate at which they are cooled, as lower cooling rates lead to materials that lie deeper in the underlying potential energy landscape. 
\par To simulate two-dimensional liquids and liquid-cooled glasses, we used the Kob-Andersen model, which consists of a binary mixture of spheres whose glass-forming behavior in the bulk has been studied extensively. \cite{K} This binary mixture is comprised of 65\% type A and 35\% type B particles which have unit mass and which interact according to the pairwise Lennard-Jones potential,
\begin{equation}\nonumber
V_{ij}(r) = \begin{cases} 4\epsilon_{ij}\left[\left(\frac{\sigma_{ij}}{r}\right)^{12} - \left(\frac{\sigma_{ij}}{r}\right)^{6}\right] & r \le r_{cut, ij} \\ 0 & r > r_{cut, ij} \end{cases} \hspace{5mm} i, j \in \{A, B\},
\end{equation}
where $r$ is the distance between a pair of particles, $\epsilon_{ij}$ characterizes the depth of the potential, and $\sigma_{ij}$ characterizes the finite distance below $r_{cut, ij}$ at which the potential is zero. Specific parameter values for different values of $i$ and $j$, given in Lennard-Jones units, are shown in Table \ref{tab:table1}.

\begin{table}
\caption{\label{tab:table1}\small \indent Lennard-Jones parameters for simulations of liquids and liquid-cooled glasses. Note that $\epsilon_{ij}=\epsilon_{ji}$, $\sigma_{ij} = \sigma_{ji}$, and $r_{cut, ij}= r_{cut, ji}$.}
\begin{ruledtabular}
\begin{tabular}{ccc}
$\epsilon$ & $\sigma$ & $r_{cut}$\\
\hline
$\epsilon_{AA} = 1.00$ & $\sigma_{AA} = 1.00$ & $r_{cut, \hspace{0.5mm} AA} = 2.50$ \\
$\epsilon_{AB} = 1.50$ & $\sigma_{AB} = 0.80$ & $r_{cut, \hspace{0.5mm} AB} = 2.00$\\
$\epsilon_{BB} = 0.50$ & $\sigma_{BB} = 0.88$ & $r_{cut, \hspace{0.5mm} BB} = 2.20$\\
\end{tabular}
\end{ruledtabular}
\end{table}

\par We performed 20,000 independent simulations of this model using LAMMPS. \footnote{See https://lammps.sandia.gov/} Each simulation contained a fixed total of 4,320 particles in a simulation box of length $60 \sigma_{AA}$ in the x- and y-directions.\footnote{We chose this system size to be consistent with the simulations in Reid \textit{et al.} \cite{R} The units used in these simulations are Lennard-Jones units.} Each simulation was performed in the canonical NVT ensemble with periodic boundaries in the x- and y-directions, and the temperature was reduced linearly from an initial temperature of 2.0 to a final temperature of 0.05 in Lennard-Jones units using a Nose-Hoover thermostat for $t_{cool}$ simulation steps. During each simulation we recorded the inherent structure energy per particle of the system using the FIRE minimization algorithm. \cite{F} The inherent structure energy of a configuration, used to quantify a configuration's stability, is the potential energy brought to its local minimum. \cite{R} 
\par Each of these simulations models a liquid that cools and solidifies into a glass below $T_g$. For 10,000 of the simulations we used $t_{cool} = 2\times 10^7$ cooling steps, and for the other 10,000 we used $t_{cool}=2\times 10^5$ cooling steps. We calculated the average glass transition temperature, $T_g$, for each cooling rate by identifying two linear regimes in each of the average inherent structure energy curves, one corresponding to the supercooled liquid regime and the other to the glass regime. \cite{JDP1} We then fit the data in these regimes and calculated the intersection point, as shown in Figure \ref{fig:figure1}(a), which gives $T_g \approx 0.37$ for $t_{cool} = 2\times 10^7$ and $T_g \approx 0.39$ for $t_{cool} = 2\times 10^5$. The glass configurations generated using $t_{cool} = 2\times 10^7$ are at a lower average inherent structure energy than those generated using $t_{cool} = 2\times 10^7$ because $t_{cool} = 2\times 10^7$ corresponds to a lower cooling rate. 

\subsection{Datasets} Our goal was to train machine learning algorithms to perform a binary classification task: identify a particle configuration as a liquid or a glass. To that end, we used the particle configurations generated by the simulations described in \S\ref{sec:sec21} to construct six datasets. 
\par
\begin{figure*}
\centering
\hspace{-5mm}
\subfloat[]{
  \includegraphics[scale = 0.5]{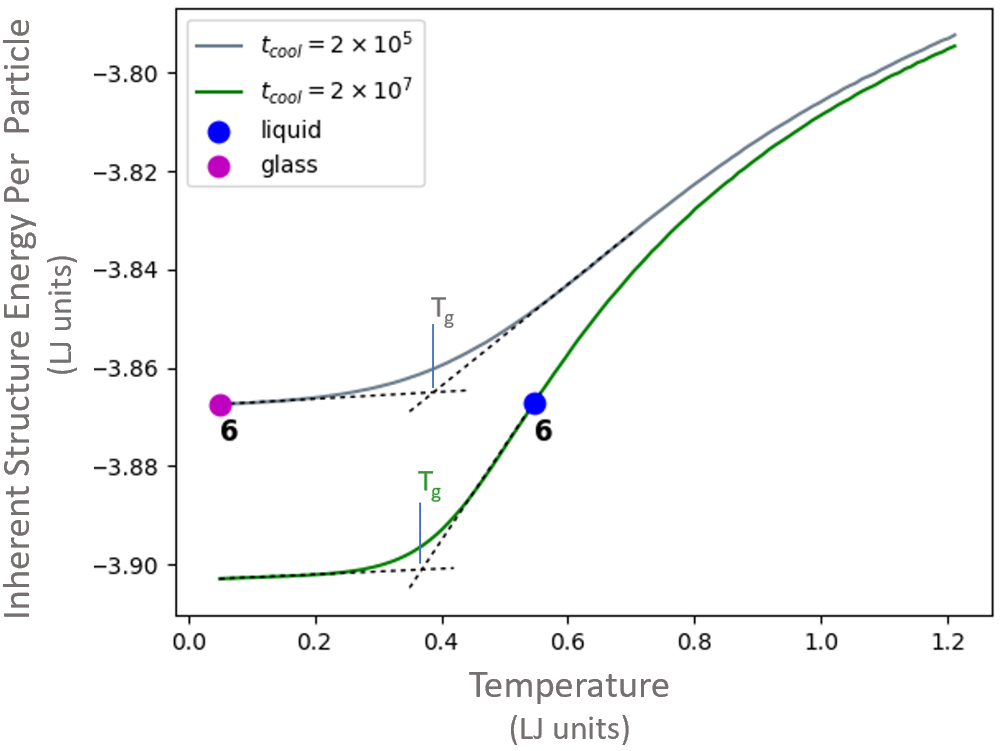}}%
  \hspace{20mm}
\subfloat[]{
  \includegraphics[scale = 0.52]{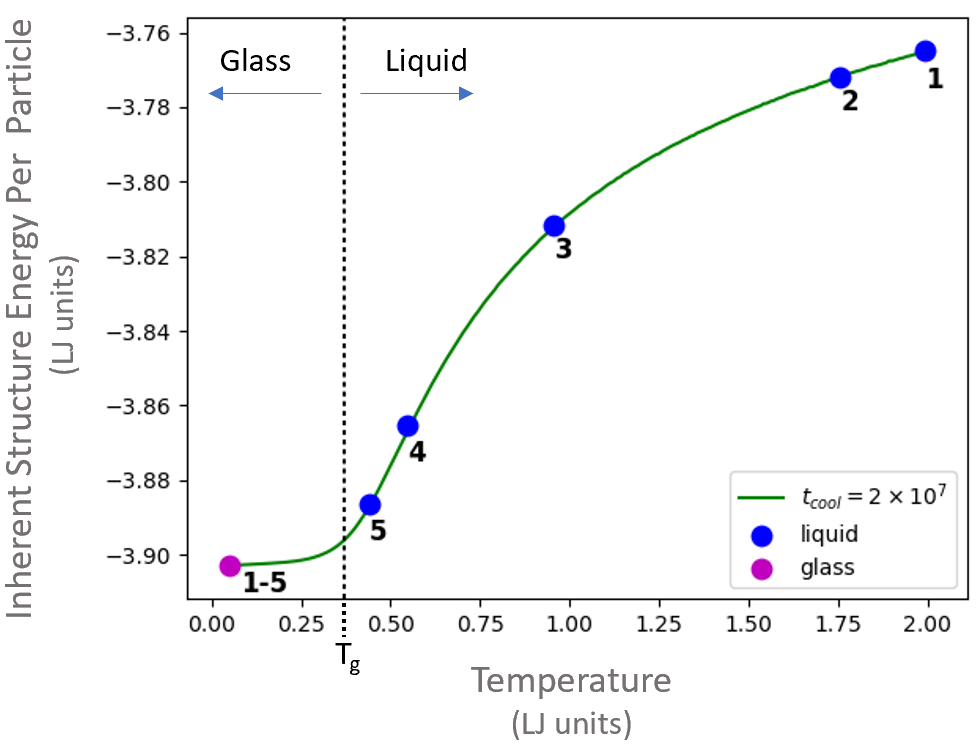}}%
\caption{\label{fig:figure1}\small \textbf{(a)} Average inherent structure energies per particle as a function of temperature for 10,000 configurations with $t_{cool} = 2\times 10^5$ and 10,000 configurations with $t_{cool} = 2\times 10^7$. $T_g$ for each curve is calculated as the intersection of linear approximations to the supercooled liquid regime and the glass regime, as shown by the dotted black lines. We consider configurations at temperatures above $T_g$ to be liquids and configurations below $T_g$ to be glasses. The magenta and blue points indicate the temperatures at which we select glass and liquid configurations for dataset 6, respectively. Standard deviations of energy values, not shown here, are on the order of $10^{-3}$. \textbf{(b)} Average inherent structure energy per particle as a function of temperature for the 10,000 configurations with $t_{cool} = 2\times 10^7$. The magenta point indicates the temperature at which we select glass configurations for datasets 1 through 5. The blue points indicate the temperatures at which we select liquid configurations, labeled by dataset number.}
\end{figure*}

 Datasets 1 through 5 are each composed of 10,000 glass configurations and 10,000 liquid configurations, all taken from the simulations with $t_{cool} = 2\times 10^7$. The glass configurations in each dataset are all at a temperature of 0.05. In order of increasing dataset number, the liquid configurations are at temperatures 1.99, 1.76, 0.96, 0.55, and 0.44, as enumerated in Figure \ref{fig:figure1}(b). 
\par In datasets 1 through 5, the liquid configurations are at higher inherent structure energies than the glass configurations. In order to compare liquid and glass configurations at the same average inherent structure energy, we constructed dataset 6, labeled in Figure \ref{fig:figure1}(a). This dataset has 10,000 glass configurations at a temperature of 0.05 taken from simulations with $t_{cool} = 2\times 10^5$ and 10,000 liquid configurations at a temperature of 0.55 taken from simulations with $t_{cool}=2\times 10^7$.
\par Each dataset serves as a test for a machine learning algorithm's ability to distinguish between amorphous material structures, and they are numbered roughly in order of increasing difficulty. Dataset 1 represents a relatively easier test, because the liquid and glass configurations in this dataset have a large difference in average inherent structure energy and distinctly different local structures, as exhibited by their average radial distribution functions (see Figure \ref{fig:figure11} in the Supplementary Information). Datasets 2 through 5 represent increasingly difficult tests as the differences in average inherent structure energy decrease and the radial distribution functions converge. Dataset 6 represents one of the most difficult tests, because the liquid and glass configurations are, on average, at the same inherent structure energy, as shown in Figure \ref{fig:figure1}(a), and have very similar average radial distribution functions (see Figure \ref{fig:figure12} in the Supplementary Information). 

\subsection{Convolutional Neural Networks} In this section we give a brief overview of convolutional neural networks (CNNs) and a description of the architecture and training routine for CNNs used in this work.
\par Neural networks have been proven capable of approximating a wide set of functions. \cite{U} Convolutional neural networks, which use convolutional layers in addition to the fully connected layers that characterize a basic feed forward network, have been shown to excel at computer vision tasks ranging from assessing cancer risk in radiology scans to galaxy morphology classification in telescope images. \cite{I, BC, GMC} As explained in \S\ref{sec:sec1}, CNNs have also been used to classify a variety of materials, including crystal structures and Ising model configurations.
\par In a convolutional layer, a set of matrices, or kernels, is convolved with an input matrix to produce a set of output matrices, or feature maps. \cite{G} Notably, convolutions are equivariant to translation, a property that underlies the effectiveness of CNNs in detecting features such as edges and shapes in different locations of an image, while also significantly reducing the number of parameters compared to a basic fully connected feedforward network. 

\subsubsection{Network Architecture} 
\begin{figure}
\centering
\includegraphics[scale = 0.5]{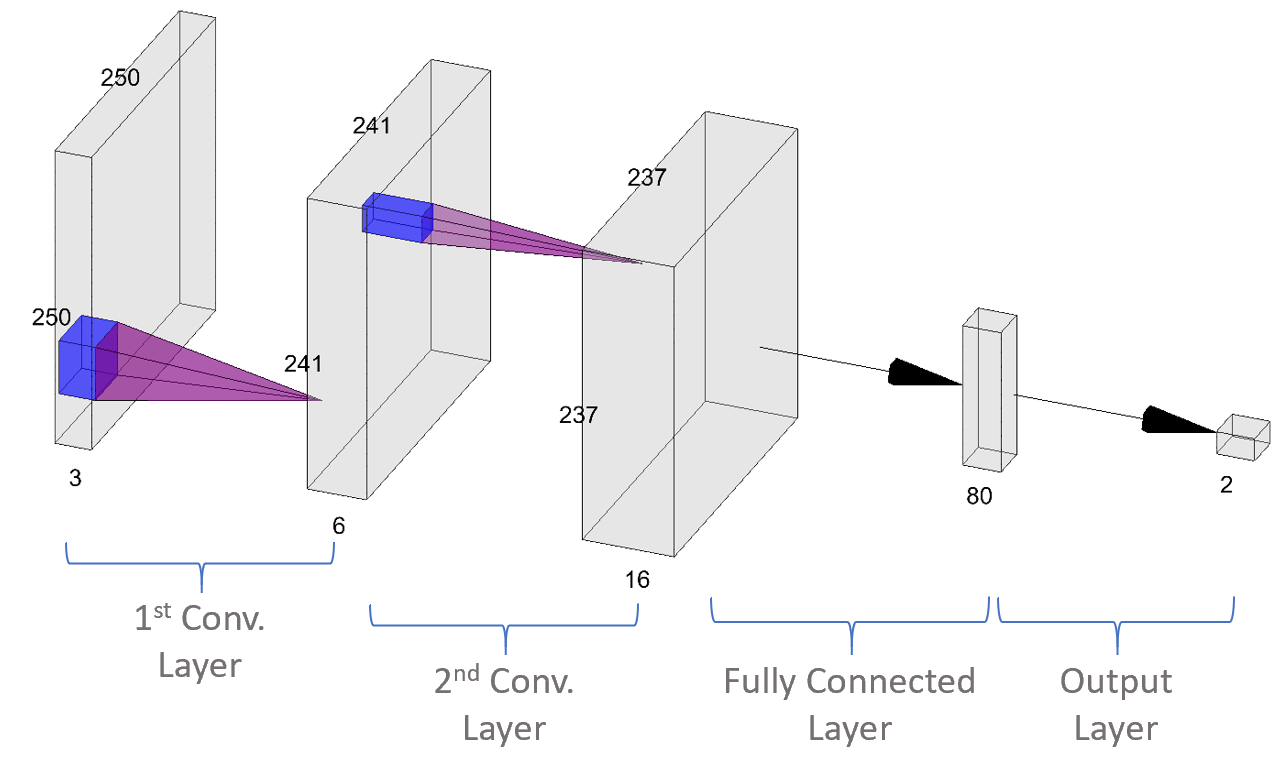}
\caption{\label{fig:figure2} \small Convolutional neural network (CNN) architecture used in this work. Two convolutional layers are followed by a fully connected layer and an output layer. In order to be analyzed by a CNN, particle configurations are rendered as images.}
\end{figure}

The CNN that we developed, shown in Figure \ref{fig:figure2}, has two convolutional layers. The first has 6 kernels with dimensions $10 \times 10 \times 3$ along with a bias vector and ReLU activation, with no zero padding and a stride of 1. The second has 16 kernels with dimensions $5 \times 5 \times 6$ along with a bias vector and ReLU activation, also with no zero padding and a stride of 1. The 16 feature maps that are output by these layers, which have dimensions $237 \times 237$, are then flattened and fed into a fully connected layer with 80 neurons and a bias vector, followed by ReLU activation. Dropout, a regularization method that has been shown to reduce overfitting,\cite{JMLR:v15:srivastava14a} is applied to this fully connected layer followed by a final output layer with two neurons. We used TensorFlow to build and train these models, \footnote{See www.tensorflow.org.} and all of our code is available in our GitHub repository.\footnote{See https://github.com/ks8/glassML.}

\subsubsection{Training}\label{sec:sec232} 
\begin{figure}
\centering
\includegraphics[scale = 0.5]{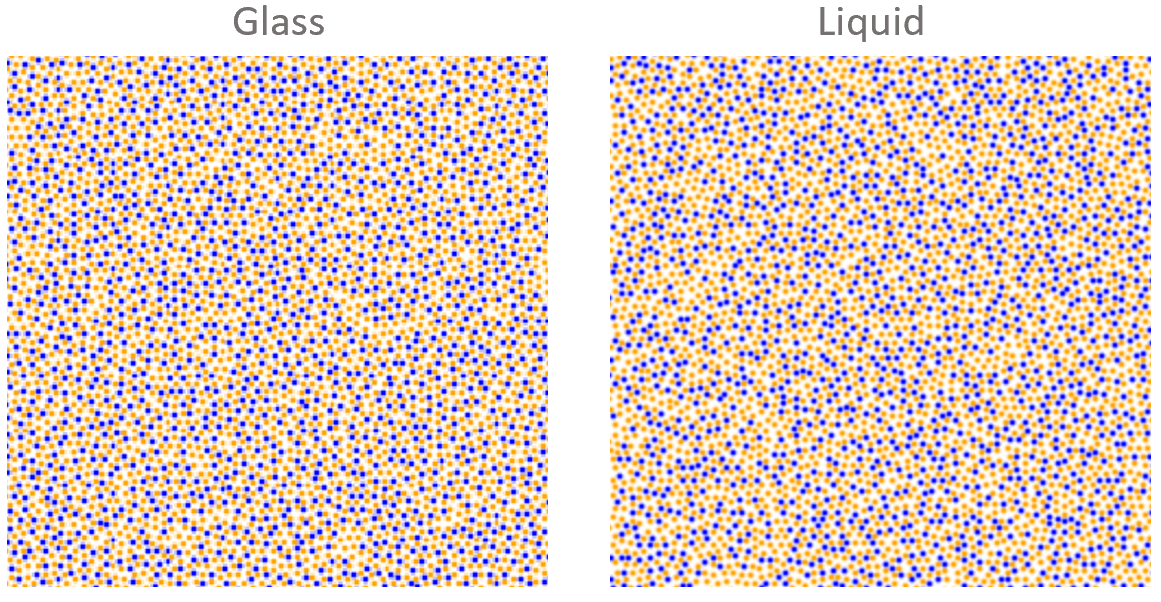}
\caption{\label{fig:figure3} \small Representative images of glass and liquid particle configurations from dataset 1 prepared for a CNN. Orange dots are type A particles and blue dots are Type B particles.}
\end{figure}

To prepare a dataset for a CNN, we rendered each particle configuration as a 250 x 250 pixel PNG image, as shown in Figure \ref{fig:figure3}.\footnote{This diagram was produced using the tools at http://alexlenail.me/NN-SVG/LeNet.html.} During training we use an on-the-fly data augmentation scheme whereby images are uniformly randomly rotated by 0, 90, 180, or 270 degrees and then flipped across the $y = 0.5$ axis (upside down) with a probability of 0.5, which effectively expands our dataset size. Each batch array is a tensor with dimensions $n_b \times 250 \times 250 \times 3$, where $n_b$ is the batch size and the last dimension corresponds to the three color channels (red, green, blue) for color images.   
\par Backpropagation is performed using cross-entropy loss with $L^2$ regularization (coefficient of 0.01) on all four layers of the network. Weights are updated using the ADAM optimization procedure. \cite{AO} We used a learning rate schedule with a piecewise linear increase and exponential decay: the learning rate increases linearly from an initial learning rate of $1\times 10^{-4}$ to a maximum learning rate of $1\times 10^{-3}$ during the first two epochs, and then it decreases exponentially to a learning rate of $1\times 10^{-4}$ during the remaining epochs. \cite{A}

\subsection{Message Passing Neural Networks} In this section we give a brief overview of message passing neural networks (MPNNs) and a description of the architecture, training routine, and interpretation scheme for MPNNs used in this work.

\par As explained in \S\ref{sec:sec1}, MPNNs have been used to predict properties of molecules and crystals by representing them as graphs, with atoms corresponding to nodes and bonds corresponding to edges. To the best of our knowledge, however, MPNNs have never before been used to classify and analyze large ensembles of particles or amorphous materials such as glasses. 

\par As described in the message passing framework established by Gilmer \textit{et al.}, MPNNs operate on undirected graphs $G$ with node features $x_v$ and edge features $e_{vw}$. \cite{GG} The MPNN processes these graphs in two phases: a message passing phase and a readout phase. In the message passing phase, the MPNN builds a representation of the input graph, and in the readout phase, the MPNN uses this representation to predict properties of interest. 

\par The message passing phase runs for $\mathcal{T}$ steps. During each step $t$, hidden states $h_v^t$ and messages $m_v^{t+1}$ at each node $v$ in the graph are updated using message function $M_t$ and vertex update function $U_t$ according to

\begin{align}
\begin{split}
m_v^{t+1} &= \sum_{w\in N(v)}M_t(h_v^t, h_w^t, e_{vw}) \\
h_v^{t+1} &= U_t(h_v^t, m_v^{t+1}),
\end{split}
\end{align}
where $N(v)$ are the neighbors of $v$ in graph $G$ and $h_v^0$ is a function of the initial node features $x_v$. The readout phase uses some function $R$ to make a property prediction of interest based on the final hidden states according to

\begin{equation}
\hat{y} = R\left(\left\{h_v^\mathcal{T}\right|v\in G\}\right).
\end{equation}

\subsubsection{Network Architecture} Here we describe the particular variant of MPNN that we use in this work, as described in Yang \textit{et al.} and proposed in Dai \textit{et al.} \cite{S, DD} This variant, called Directed MPNN (D-MPNN), uses messages associated with directed edges rather than messages associated with nodes. Our motivation in using this particular architecture is that having messages passed along edges instead of nodes in the graph is more conducive to extracting an interpretation of the network using self-attention, as described in \S\ref{sec:sec243}.
\par
\begin{figure}
\centering
\includegraphics[scale = 0.9]{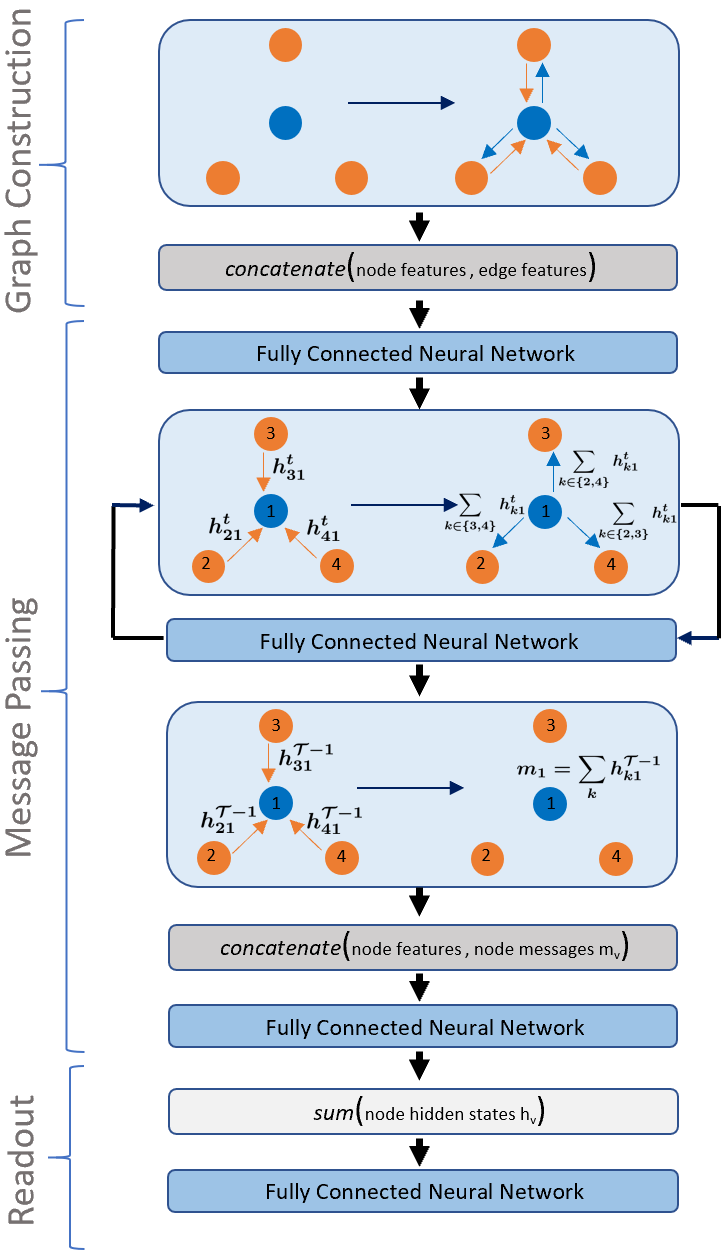}
\caption{\label{fig:figure4}\small Directed message passing neural network (D-MPNN) architecture used in this work. First, particle configurations are transformed into a graph representation by connecting each particle to its $k$ nearest neighbors (graph construction). Then, after the edge hidden states are initialized via a fully connected layer, message passing along the edges is carried out, as in Eqns. \ref{eqn:eqn4} - \ref{eqn:eqn8} (message passing phase). Finally, property predictions are generated via a fully connected neural network, as in Eqns. \ref{eqn:eqn9} and \ref{eqn:eqn10} (readout phase).}
\end{figure}

 The D-MPNN works as follows, as shown in Figure \ref{fig:figure4}. It operates on hidden states $h_{vw}^t$ and messages $m_{vw}^t$ at each edge connecting nodes $v$ and $w$. The edges in graph $G$ are directed, so that for any two connected nodes $v$ and $w$, there is an edge from $v$ to $w$ and an edge from $w$ to $v$. In this way, the messages are also directed: $h_{vw}^t$ and $m_{vw}^t$ are distinct from $h_{wv}^t$ and $m_{wv}^t$. In the message passing phase, the hidden states and messages are updated according to 

\begin{align}
\begin{split}\label{eqn:eqn4}
m_{vw}^{t+1} &= \sum_{k\in \{N(v) \backslash w\}}M_t(x_v, x_k, h_{kv}^t) \\
h_{vw}^{t+1} &= U_t(h_{vw}^t, m_{vw}^{t+1}).
\end{split}
\end{align}
Note that the message $m_{vw}^{t+1}$ is not a function of the reverse message $m_{wv}^t$ from the previous step. Prior to the first step of message passing, edge hidden states are initialized according to 

\begin{align}
\begin{split}
h_{vw}^0 &= \alpha\left(W_i\hspace{1mm}cat(x_v, e_{vw})\right), 
\end{split}
\end{align} 
where $W_i\in \mathbb{R}^{h\times h_i}$ is a learned matrix, $cat(x_v, e_{vw})\in\mathbb{R}^{h_i}$ is the concatenation of the node features $x_v$ for node $v$ and the edge features $e_{vw}$ for edge $vw$, and $\alpha$ is the ReLU activation function. The message passing functions $M_t$ are given by 

\begin{align}
\begin{split}
M_t(x_v, x_w, h_{vw}^t) = h_{vw}^t,
\end{split}
\end{align}
and the edge update functions are given by a neural network,

\begin{align}
\begin{split}
U_t(h_{vw}^t, m_{vw}^t) = U(h_{vw}^t, m_{vw}^t) = \alpha\left(h_{vw}^0 + W_mm_{wv}^{t+1}\right),
\end{split}
\end{align}
where $W_m\in\mathbb{R}^{h\times h}$ is a learned matrix with hidden size $h$. Dropout is then applied. The presence of $h_{vw}^0$ in the above equation provides a skip connection to the original feature vector for that edge. 
After $\mathcal{T}-1$ steps of this message passing, a node representation of the graph is constructed by summing inbound edge features in a final message passing step according to 

\begin{align}
\begin{split}\label{eqn:eqn8}
m_v &= \sum_{k\in N(v)}h_{kv}^{\mathcal{T} -1} \\
h_v &= \alpha\left(W_a\hspace{1mm} cat(x_v, m_v)\right),
\end{split}
\end{align}
where $W_a\in\mathbb{R}^{h\times h}$ is a learned matrix. 
\par The readout phase is the same as for general MPNNs. For the readout function $R$, node hidden states are summed to obtain a feature vector for the graph

\begin{align}
\begin{split}\label{eqn:eqn9}
h = \sum_{v\in G}h_v.
\end{split}
\end{align} 
Property predictions are then generated according to
\begin{align}
\begin{split}\label{eqn:eqn10}
\hat{y} &= f(h),
\end{split}
\end{align}
where $f$ is a feed forward neural network with ReLU activation and dropout at each layer.

\subsubsection{Training}\label{sec:sec242} To prepare a dataset for a D-MPNN we extract a matrix of scaled particle coordinates $x, y \in [0, 1]$ and particle types (A or B) for each configuration. In order to be processed by a D-MPNN, the raw coordinate data for a batch of configurations is transformed into a graph, which happens on-the-fly for each batch that is loaded for training. 
\par To do this, we first implement a step for each configuration in a batch that effectively increases the size of the dataset. We specify a hyperparameter called the window length, $l_{win} \in (0, 1)$, whose square is the area of a square window that we randomly select from each particle configuration. We select this window by drawing two random numbers uniformly from the interval [0, 1 - $l_{win}$], called $x_{rand}$ and $y_{rand}$, and then select particles whose $x, y$ coordinates satisfy $x_{rand} \le x \le x_{rand} + l_{win}$ and $y_{rand} \le y \le y_{rand} + l_{win}$. For example, if $l_{win} = 0.5$, we select a subset of particles in a random square window corresponding to about 25\% of the original set of particles. This process effectively augments the size of the dataset.   
\par We then transform this batch of modified configurations into a graph. Each individual configuration is transformed into a graph by connecting each particle to its $k$ nearest neighbors, as illustrated in Figure \ref{fig:figure4}. Each connection is comprised of two edges, one directed from particle (node) $v$ to particle (node) $w$, and the other in the reverse direction. For each edge we compute the Euclidean distance between the connected nodes. Thus, the node features $x_v$ in this graph are the $x$-coordinate, $y$-coordinate, and particle type, $t$, for node $v$, and the edge feature $e_{vw}$ is the distance between nodes $v$ and $w$. The graphs corresponding to individual configurations in a batch are then concatenated into a single larger graph representing the entire batch. In this batch graph, the nodes corresponding to one configuration are connected to each other but not to nodes corresponding to other configurations. 
\par During training, backpropagation is performed using cross-entropy loss. Weights are updated using the ADAM optimization procedure, \cite{AO} and we used the same learning rate schedule described in \S\ref{sec:sec232}. Our code for D-MPNNs, written in PyTorch, \footnote{See https://pytorch.org/.} is built upon publicly available code. \footnote{See https://github.com/rusty1s/pytorch\_geometric and https://github.com/wengong-jin/chemprop.} All of our code is available in our GitHub repository.\footnote{See https://github.com/ks8/glassML.}

\subsubsection{Interpretation}\label{sec:sec243} Currently, interpreting a neural network is very challenging. However, new techniques are beginning to provide avenues for accomplishing this task. One such technique is called self-attention, which is essentially a mechanism that allows us to examine which features of the data a neural network is paying "attention" to most.\cite{A, S2} Mathematically, self-attention is akin to a dot product that yields a set of weights for each feature, which are then interpreted as "attention" scores.

\par We place a self-attention mechanism on the edges of a graph and apply it during the message passing phase for the first $\mathcal{T} - 1$ steps. This allows us to examine which edges of a graph the D-MPNN is paying the most "attention" to while training, which could give us insight into how the network is making its classification decisions. 
\par Our self-attention mechanism for D-MPNNs works as follows. According to Eqn. \ref{eqn:eqn4}, in round $t + 1$ of message passing, the feature vector for each edge $e_{vw}$ in a graph is updated to $h_{vw}^{t + 1}$. Each of these vectors has length $h$ (hidden size), and we assume that there are $n$ edges in the graph. After Eqn. \ref{eqn:eqn4}, we insert the following steps into the D-MPNN algorithm. The feature vectors $h_{vw}^{t + 1}$ are concatenated as row vectors in a matrix $\textbf{h}$, where $\textbf{h} \in \mathbb{R}^{n\times h}$. We then apply the following transformation to $\textbf{h}$: 
\begin{align}
\begin{split}
\boldsymbol\sigma = \text{softmax}\left(\alpha\left(\textbf{h}^T\textbf{W}_{attn}\right)\textbf{v}_{attn}\right),
\end{split}
\end{align}
where $\alpha$ is the ReLU transformation, $\textbf{W}_{attn}\in \mathbb{R}^{h \times h}$ is a learned parameter matrix, and $\textbf{v}_{attn}\in \mathbb{R}^{h}$ is a learned parameter vector. The weights $\boldsymbol\sigma$ are then dotted with the initial row vectors to yield a new set of hidden edge states:
\begin{align}
\begin{split}
\textbf{h}^{\prime} = \boldsymbol\sigma\textbf{h}.
\end{split}
\end{align}
Finally, the row vectors of $\textbf{h}^\prime$ replace the corresponding values of the edge features from Eqn. \ref{eqn:eqn4} and are applied as such to the next round of message passing. The weights $\boldsymbol{\sigma}$ represent the "attention" that the network is giving to each edge in the graph. As discussed in \S\ref{sec:sec3}, we then used the \textit{networkx} package in Python to quantify attributes of the graph structure of these self-attention weights for glass and liquid particle configurations. 

\subsection{Hyperparameter Optimization and Cross Validation}\label{sec:sec25}
There are seven hyperparameters in our D-MPNN: window length $l_{win}$, number of nearest neighbors $k$, number of message passing steps $\mathcal{T}$, dropout probability $p$, hidden size $h$, number of feed forward layers $f$, and batch size $n_b$. There are two hyperparameters in our CNN: dropout probability and batch size. To discover optimal values for these hyperparameters, we used a Bayesian optimization scheme called a Tree-structured Parzen Estimator (TPE), implemented in the \textit{hyperopt} package. \footnote{See https://hyperopt.github.io/hyperopt} As described in Bergstra \textit{et al.}, TPE is a type of sequential model-based global optimization algorithm which discovers optimal hyperparameters by modeling the loss function with a surrogate probability model and making increasingly well-informed guesses for a specified number of iterations (see Supplementary Information for more details).\cite{B} These algorithms have been shown to exceed the performance of grid search and random search when optimizing for multiple hyperparameters. Table 2 shows the range of hyperparameter values that we explored using \textit{hyperopt}. 

\begin{table}
\caption{\label{tab:table2}\small Ranges of values used for hyperparameter optimization. To discover optimal values in these ranges, we used a Bayesian optimization scheme called a Tree-structured Parzen Estimator (TPE). \cite{B} TPE discovers optimal hyperparameters by modeling the loss function with a surrogate probability model and making increasingly well-informed guesses for a specified number of iterations. We found that batch sizes greater than five for D-MPNNs sometimes exceeded GPU memory, so we fixed $n_b = 5$ for D-MPNNs.}
\begin{ruledtabular}
\begin{tabular}{cccc}
Hyperparameter & Low & High & Step Size\\
\hline
\# nearest neighbors ($k$) & 1 & 5 & 1 \\
window length ($l_{win}$) & 0.1 & 0.3 & 0.05 \\
hidden size ($h$) & 300 & 2400 & 100 \\
\# message passing steps ($\mathcal{T}$) & 2 & 6 & 1 \\
dropout probability ($p$) & 0.0 & 0.5 & 0.05 \\
\# feed forward layers ($f$)  & 1 & 3 & 1 \\
batch size ($n_b$) & 10 & 100 & 10 \\
\end{tabular}
\end{ruledtabular}
\end{table}

To measure the performance of a CNN or D-MPNN on a dataset, we used a three-fold nested cross-validation scheme, which works as follows. We select three disjoint subsets of the dataset, each containing 20\% of the data, which we call outer test sets. For a given outer test set, we label the remaining 80\% of the data as the outer train set. We then split the outer train set into an 80\% inner train set, 10\% inner validation set, and 10\% inner test set. We apply 15 iterations of the TPE algorithm, where each iteration consists of 10 epochs of training using the inner train and validation sets, to identify an optimal set of hyperparameters on the inner test set. We train a model with these optimal hyperparameters on the outer train set and report performance on the outer test set. The outer train sets consist of 80\% of a dataset, which is 16,000 configurations. Therefore, each CNN is trained on 16,000 different samples. Since each MPNN is trained on random square windows of these configurations for 10 epochs, each MPNN is trained on 160,000 different samples. This process is repeated for each of the three outer test sets, or folds, giving us three independent measures of performance for which we can report the mean and standard deviation. 
\par The primary performance metric that we report, which we also use to select optimal sets of hyperparameters, is AUC, or area under the ROC (receiver operating characteristic) curve. Our machine learning models output a continuous value between 0 and 1 when making a prediction, so a threshold value is used to binarize the prediction. For example, if the threshold value is 0.5, outputs greater than 0.5 correspond to glass while outputs less than 0.5 correspond to liquid. The ROC curve is created by plotting the true positive rate against the false positive rate at different thresholds. AUC ranges from 0 to 1, where 0.5 indicates that classification is no better than random guessing and 1 indicates that classification is perfect across all thresholds. For comparison, we also report accuracy, or fraction of configurations labelled correctly using a single default threshold.

\section{Results and Discussion}\label{sec:sec3}

\subsection{CNNs} The CNNs classified liquid and glass configurations in datasets 1 through 6 with greater than 0.98 AUC, as shown in Figure \ref{fig:figure5} (see Figure \ref{fig:figure15} in the Supplementary Information for a complementary accuracy report). These results indicate that CNNs are capable of classifying amorphous materials with no \textit{a priori} information about particle interactions and no hand-crafted descriptors of local particle environments, even when the material structures are only subtly different. The CNNs are simply given basic geometric information - particle coordinates, rendered as an image - and they are able to classify configurations very accurately.
\par Moreover, CNNs optimized on one dataset were able to make accurate classifications of configurations from other datasets, as exhibited in Figure \ref{fig:figure6}(a). All of the models generalized successfully to other datasets, but the models trained on the more difficult tasks (i.e. datasets 5 and 6) performed best across the board.
\par A central question that arises is how the CNNs are making classification decisions. Since a CNN can accurately identify liquid and glass configurations prepared at the same inherent structure energy (dataset 6), these algorithms are not simply computing energy as a means for classification. We tried several methods for interpreting the CNNs, including visualizing feature maps and examining whether specific geometrical patterns activated components of the network (see the Supplementary Information for more details). These methods have been used in previous work to facilitate interpretation of neural networks.\cite{IM1, IM3} However, none of these methods were successful at providing an interpretation. Further work is needed to answer this question. 
\par Besides the challenge of interpreting the CNNs, there were several other issues. The CNNs do not incorporate rotational invariance, a symmetry that is present in our data, and they require each input image to have the same size, limiting their flexibility. Moreover, rendering particle configurations as images introduces an artificial radius to these particles which is not present in the underlying system. These issues motivated our study of message passing neural networks. As discussed below, we found that message passing neural networks overcome all of these challenges.  

\begin{figure}
\hspace{-10mm}
\includegraphics[scale = 0.8]{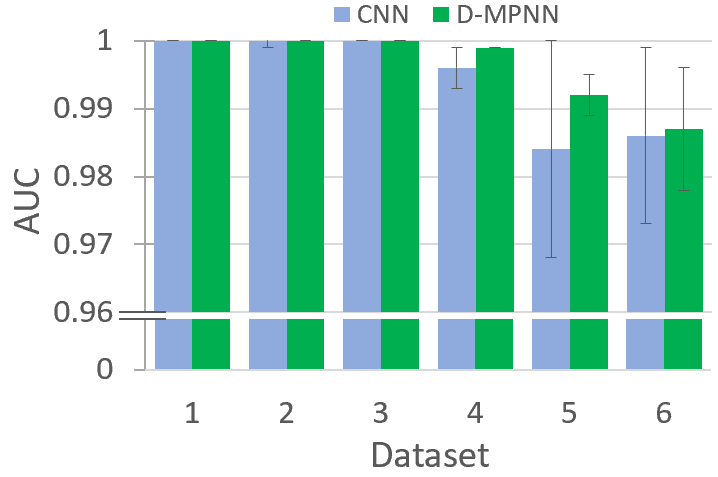}
\caption{\label{fig:figure5} \small CNN and D-MPNN average classification AUC for datasets 1 through 6 with error bars showing standard deviation. These average values were computed using the three-fold nested cross-validation scheme described in \S\ref{sec:sec25}.}
\end{figure} 

\begin{figure*}
\centering
\hspace{-5mm}
\subfloat[]{
  \includegraphics[scale = 0.4]{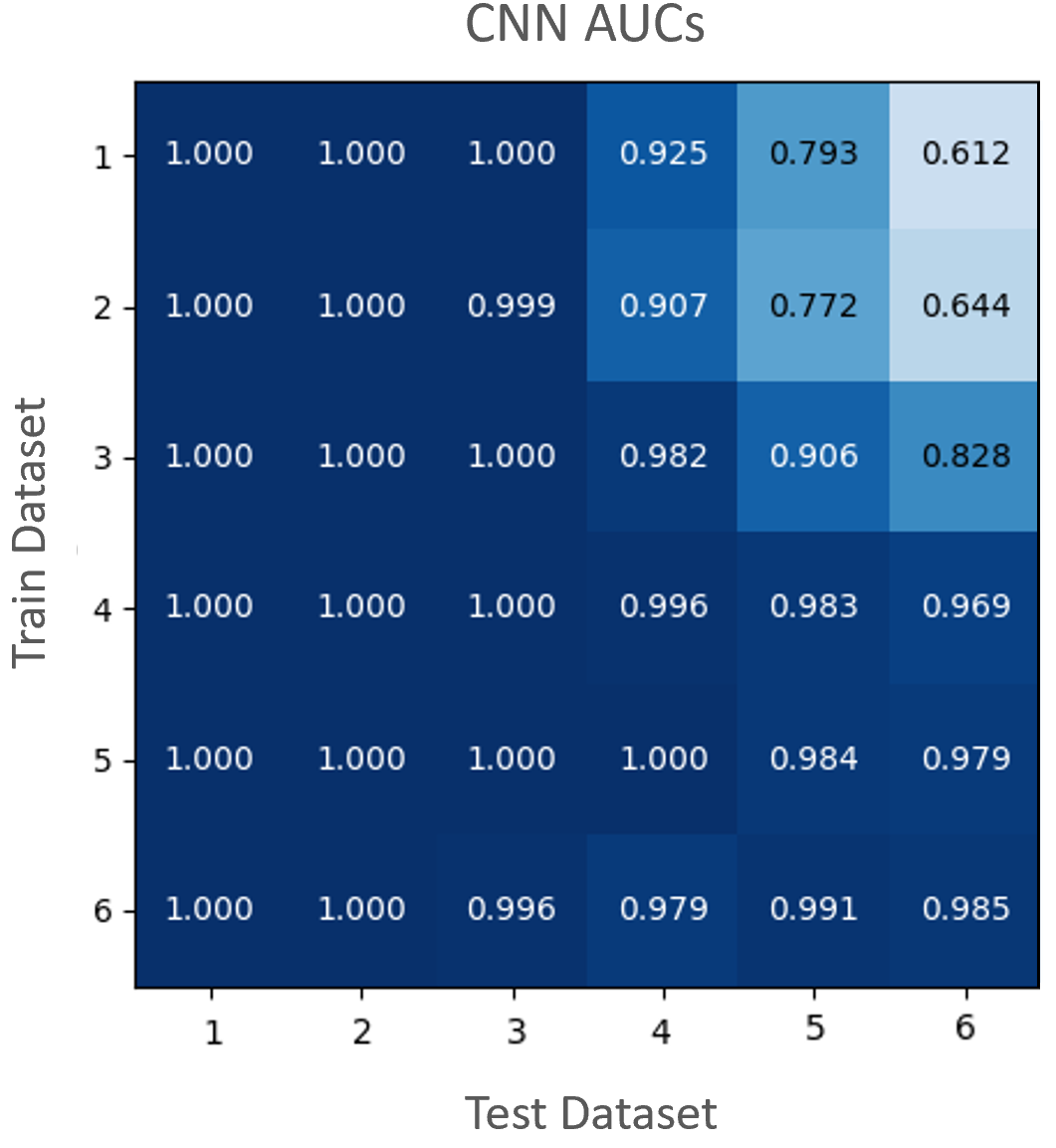}}%
  \hspace{10mm}
\subfloat[]{
  \includegraphics[scale = 0.4]{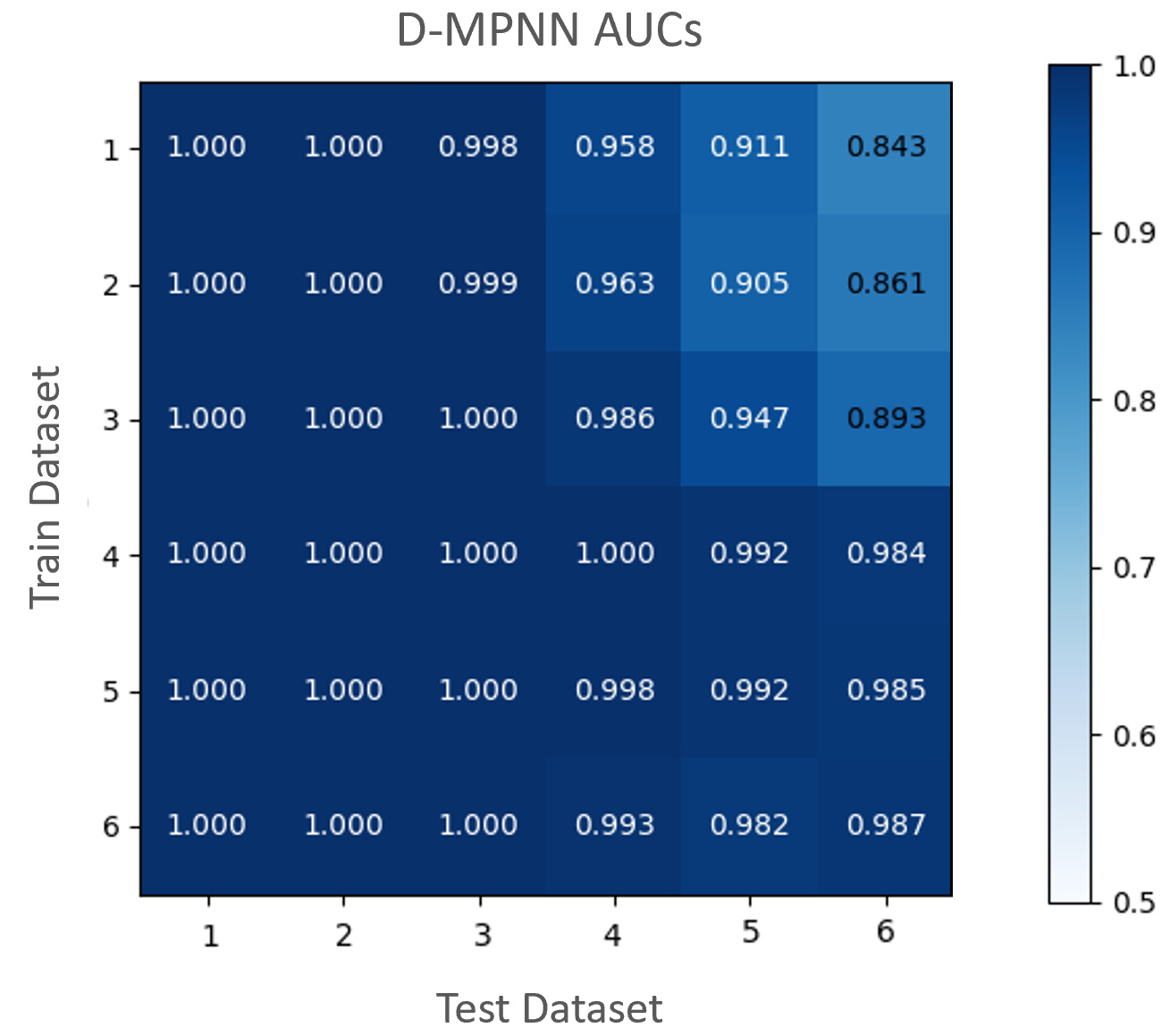}}%
\caption{\label{fig:figure6}\small These plots show how models trained on one dataset perform on all other datasets. In each of the plots in this figure, the number in row $i$, column $j$ is the average classification AUC on dataset $j$ of the optimal models trained on dataset $i$. For $i = j$ we report the average three-fold nested cross validation AUC. For $i \ne j$, we report average AUC using an outer test subset of dataset $j$. The optimal models, three CNNs and three D-MPNNs for each dataset, were generated from the three-fold nested cross validation procedure described in \S\ref{sec:sec25}.  \textbf{(a)} shows average CNN AUCs and \textbf{(b)} shows average D-MPNN AUCs.}
\end{figure*}

\subsection{D-MPNNs} D-MPNNs classified liquid and glass configurations in datasets 1 through 6 with greater than 0.98 AUC. As shown in Figure \ref{fig:figure5}, the D-MPNNs performed at least as well as the CNNs on all six datasets and had a higher average AUC than the CNNs on datasets 4 through 6 (see Figure \ref{fig:figure15} in the Supplementary Information for a complementary accuracy report). These results show that, like CNNs, D-MPNNs are capable of classifying amorphous materials with no \textit{a priori} information about particle interactions and no hand-crafted descriptors of local particle environments, even when the material structures are only subtly different. It is true that we created the particle graphs by connecting nearest neighbors, but we do not restrict the D-MPNN to focus on any specific local environments in the graph (different values for  number of nearest neighbors ($k$) are explored during optimization). 

\begin{figure}
\hspace{-10mm}
\includegraphics[scale=0.85]{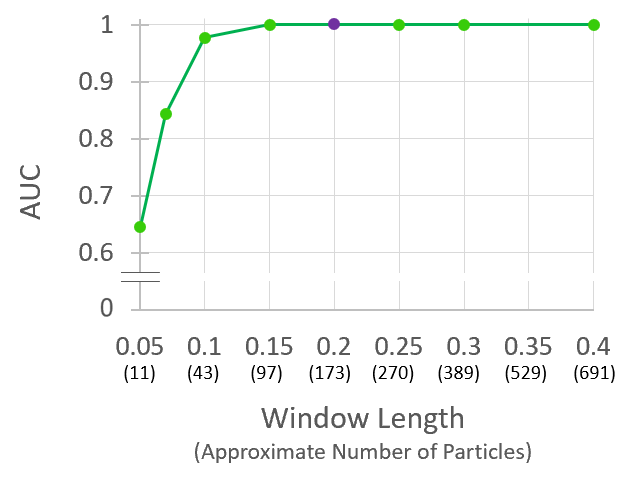}
\caption{\label{fig:figure7} \small D-MPNN AUC on an outer test set of dataset 1 with different values of $l_{win}$ (number of particles). This model was trained on dataset 1 using only $l_{win} = 0.2$ (approximately 173 particles), which is highlighted in purple.}
\end{figure} 

Similar to the CNNs, the D-MPNNs trained on one dataset were able to make accurate classifications on configurations from other datasets, as exhibited in Figure \ref{fig:figure6}(b). Again, the models trained on the more difficult tasks perform best across the board. All of the D-MPNN models, on average, generalize better than the CNN models.  
\par Besides their superior performance, the D-MPNNs have several distinct advantages over the CNNs. Because information from neighboring particles in the graph is summed, the D-MPNNs are invariant to permutations and rotations of the graph, taking advantage of a natural symmetry in the system. The D-MPNNs also operate directly on the particle coordinates without introducing any unnecessary artifacts that might appear in an image representation. In addition, they are flexible and can process different graph sizes (number of particles). We took one of the optimal D-MPNNs trained on dataset 1, which was trained using $l_{win} = 0.2$ (approximately 173 particles per graph), and successfully performed inference on both smaller and larger graphs from dataset 1, as shown in Figure \ref{fig:figure7}.

\par
\begin{figure}
\centering
\subfloat[]{
  \includegraphics[scale = 0.44]{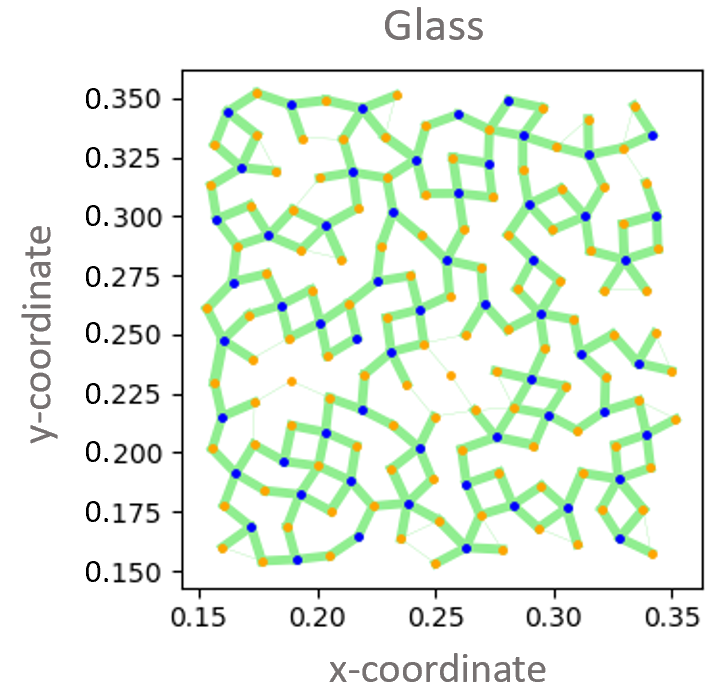}}%
\subfloat[]{
  \includegraphics[scale = 0.44]{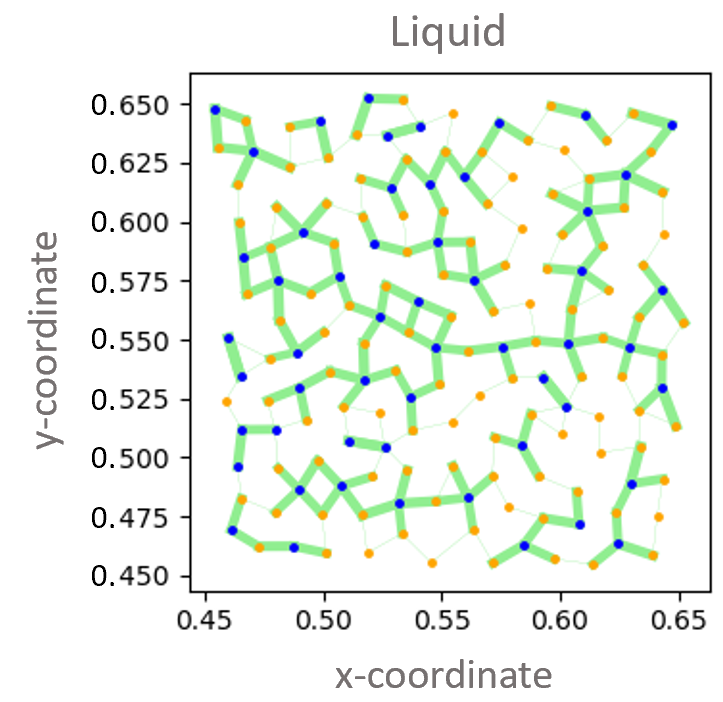}}%

\subfloat[]{
  \includegraphics[scale = 0.5]{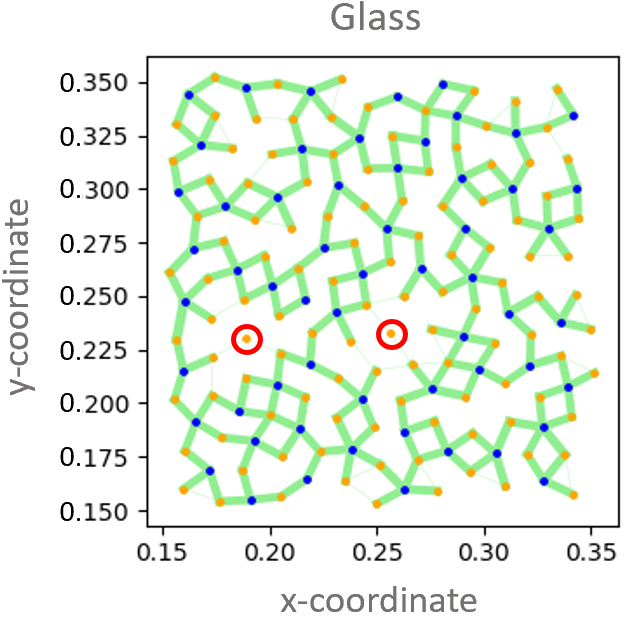}}
\subfloat[]{
  \includegraphics[scale = 0.5]{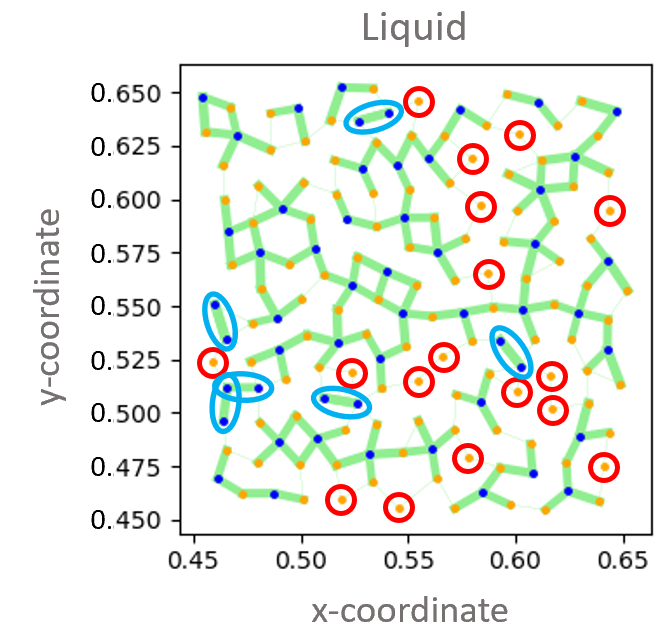}}%
\caption{\label{fig:figure8}\small Self-attention visualizations, with the attention weights computed on the $\mathcal{T} - 1^{th}$ step of message passing. All connected particles in the graph are joined with a green line whose width is proportional to the magnitude of the attention weight. Some of the edges have such small attention weights that these lines are just barely visible. Note that each connected pair of particles actually has two edges, because the graph is directed. Here, we visualize the edge with the higher weight. \textbf{(a)} Glass configuration generated with $l_{win} = 0.2$. \textbf{(b)} Liquid configuration generated with $l_{win} = 0.2$. \textbf{(c)} and \textbf{(d)} show the same attention weight visualizations as \textbf{(a)} and \textbf{(b)}, respectively, but with isolated type A particles highlighted with red circles and pairs of type B particles connected by high attention edges highlighted with blue ellipses.}
\end{figure}

 Perhaps the most significant advantage of D-MPNNs is that, when imbued with the self-attention mechanism described in \S\ref{sec:sec243}, they are able to produce a clear interpretation of how they are making classification decisions. We illustrated this by training a D-MPNN on one of the outer train sets of dataset 1 with the self-attention mechanism in place and using an optimal set of hyperparameters. The model achieved an AUC of 0.995 on the corresponding outer test set. We then visualized the attention weights from configurations in the outer test set. Representative visualizations for glass and liquid configurations are shown in Figures \ref{fig:figure8}(a) and \ref{fig:figure8}(b), respectively. 

There are clear differences in the graph structure of self-attention weights in these visualizations. The high attention weight edges form one large connected graph in the glass but multiple smaller disjoint graphs in the liquid. Several type B (blue) particles are connected to each other with high attention edges in the liquid, but none are connected in the glass. There are a larger number of type A (orange) particles that are isolated from high attention weight graphs in the liquid compared to the glass. Isolated type A particles and pairs of type B particles connected with high attention edges are highlighted in Figures \ref{fig:figure8}(c) and \ref{fig:figure8}(d). Also, all of the high attention edges in both liquid and glass are connected to at least one type B particle (there appear to be no isolated type B particles).
\par We also tested this model, which was trained using $l_{win} = 0.2$ (approximately 173 particles per graph), on larger configurations generated with $l_{win} = 0.3$ (approximately 389 particles per graph). The model successfully generalized to these larger configurations, achieving an AUC of 0.957, and yielded similar self-attention features, as shown in Figure \ref{fig:figure16} in the Supplementary Information. 

\par
\begin{figure}
\includegraphics[scale = 0.65]{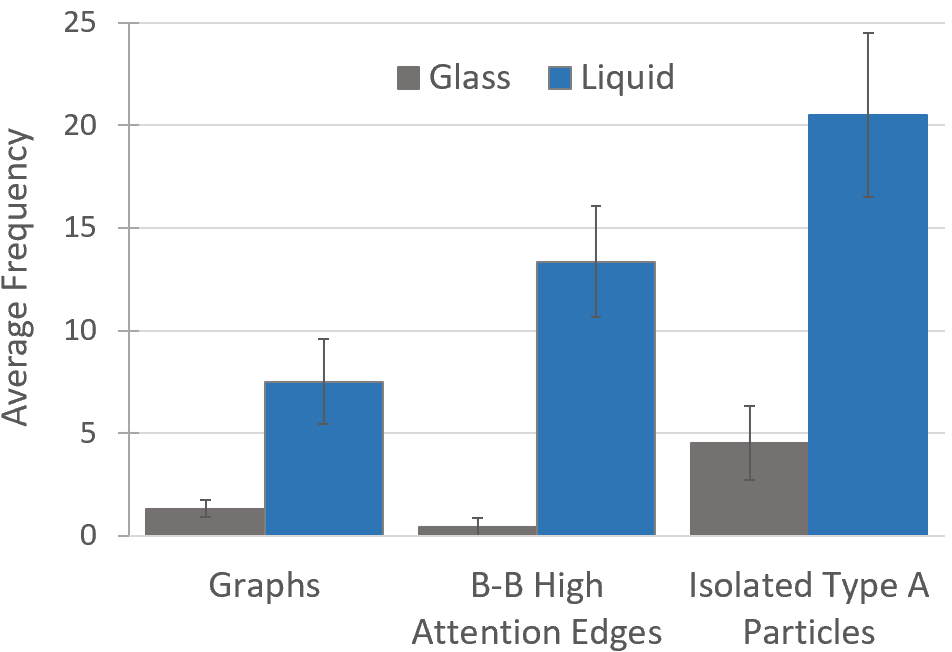}
\caption{\label{fig:figure9} \small Average number of disjoint graphs, B-B high attention edges, and isolated type A particles in an outer test set of dataset 1. High attention edges were determined with a hard cutoff. Here, a disjoint graph is a subset of the particles in a configuration connected by high attention edges. B-B high attention edges are edges connecting pairs of type B particles. Isolated type A particles are those not connected to a high attention edge.}
\end{figure}

 We quantified these features by computing the average number of disjoint graphs, average number of high attention edges connecting pairs of type B particles, and average number of isolated type A particles in dataset 1 configurations generated with $l_{win}=0.3$, as shown in Figure \ref{fig:figure9}. We confirmed that the differences in these values for glasses and liquids are statistically significant by using one-sided Wilcoxon signed-rank tests,\footnote{We used the function \textit{wilcoxon} in the Python \textit{scipy} package.} which returned p-values less than $1\times 10^{-4}$ .The number of isolated type B particles, not shown in the figure, was approximately 0 for all configurations, confirming that type B particles are nearly always adjoined to high attention edges. These features provide a clear interpretation of how the D-MPNN is making classification decisions: the network is focusing on type B particles and their relationship to nearby neighbors.    

\par Not only can we clearly interpret the D-MPNN -- a task that is typically very challenging in neural networks -- but we can also derive three novel structural metrics from this interpretation that characterize glass formation. For any configuration of particles, we construct a graph by connecting every type B particle to its two nearest neighbors, and then we count the number of disjoint graphs, the number of edges connecting pairs of type B particles, and the number of isolated type A particles. We use two nearest neighbors in this procedure because $k = 2$ was used for the D-MPNN with self-attention, and we only connect type B particles to nearest neighbors because the D-MPNN with self-attention focused on edges connected to type B particles. Note that this new graph construction procedure does not necessarily yield graphs equivalent to the attention graphs generated by a D-MPNN. It is also different from the graph construction procedure described in \S\ref{sec:sec232} because only type B particles are connected to nearest neighbors. 
\par
\begin{figure}
\includegraphics[scale = 0.75]{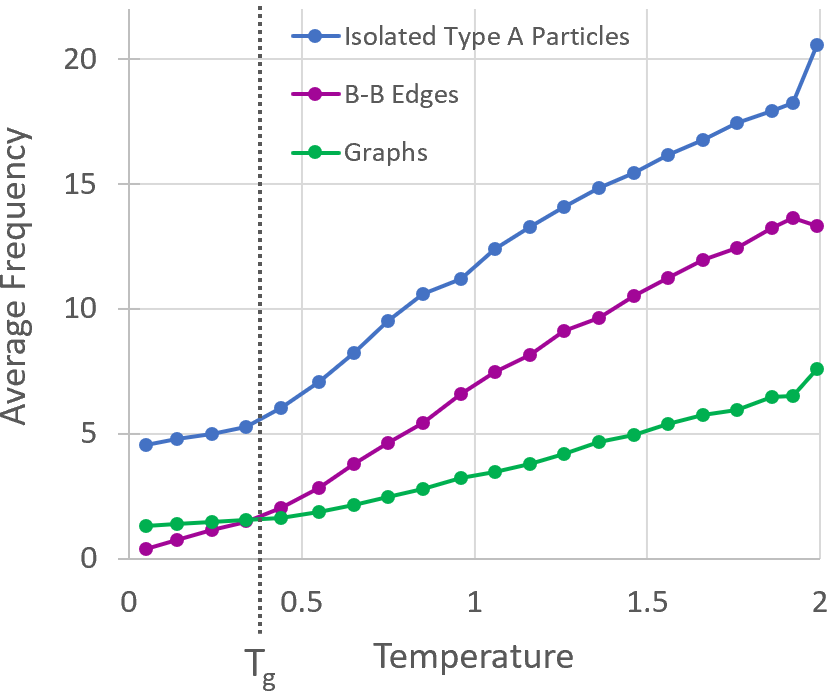}
\caption{\label{fig:figure10} \small Average number of disjoint graphs, edges connecting pairs of type B particles, and isolated type A particles in configurations from simulations with $t_{cool} = 2\times 10^7$ at a variety of temperatures. Graphs of configurations were generated by connecting every type B particle to its two nearest neighbors. Averages were computed over 1,000 configurations at each temperature.}
\end{figure}

 Average values of the three metrics are plotted as a function of temperature for configurations from simulations with $t_{cool} = 2\times 10^7$ in Figure \ref{fig:figure10}. All three of these structural metrics exhibit a similar dependence on temperature. They decrease at rapid rates above $T_g$, but immediately below $T_g$, these rates change and the metrics decrease more slowly. We fit each metric to a linear regression model of the form 
\begin{align}
\begin{split}
y_i = \alpha_0 + \alpha_1T_i + \alpha_2\mathbbm{1} + \alpha_3\mathbbm{1}T_i + \epsilon_i
\end{split}
\end{align}
using ordinary least squares, where $T$ is temperature, $\mathbbm{1}$ is an indicator variable with a value of 0 below $T_g$ and a value of 1 above $T_g$, and $\epsilon$ is a random error term. The regressions returned positive-valued interaction coefficients $\alpha_3$ with p-values less than 0.05, confirming that the difference in slopes above and below $T_g$ is statistically significant for all three metrics. Figure \ref{fig:figure17} shows these metrics plotted separately as a function of temperature along with predictions from the linear regression models. Similar to the five-fold symmetric metric in Hu \textit{et al.}, which was derived by hand, these metrics have a temperature dependence that describes the structural evolution of a liquid during glass formation. 

\par We were able to derive novel structural metrics for two-dimensional liquid and glass configurations directly from the self-attention features generated by a D-MPNN. The neural network generated these features based on raw particle coordinates alone and did not rely upon complex local descriptors, extensive spatial averaging, or a set of reference structures. This provides clear proof of concept that D-MPNNs are an effective tool not only for classifying amorphous materials but also for identifying structural features in complex systems. 

\section{Conclusion}\label{sec:sec4}
In this work, we showed that CNNs and D-MPNNs are both effective tools for amorphous materials classification, as they can classify two-dimensional liquids and liquid-cooled glasses with greater than 0.98 AUC. We also demonstrated ways in which D-MPNNs are superior to CNNs in this context, including their ability to operate on raw particle data without introducing artifacts, to achieve better classification performance, to process configurations with different numbers of particles, and to provide a clear and quantifiable interpretation of the classification process. Using the interpretation that we extracted from D-MPNNs with a self-attention mechanism, we derived three novel structural metrics that characterize glass formation. 
\par Moving forward, we believe that D-MPNNs could be applied to more difficult classification tasks that elude standard techniques. Specifically, we are interested in applying D-MPNNs to analyze other types of glassy materials, such as vapor-deposited glasses. A natural next step also includes using D-MPNNs to classify three-dimensional liquids and glasses, which could provide new insight into the local structure of these materials. In our work we performed graph-level predictions, but it is also possible to use D-MPNNs to perform node-level predictions, which could be used to extract even more granular information about local structures around specific particles or to identify defect sites in materials. And finally, further steps can be taken to improve the interpretability of attention mechanisms in these networks, including incorporating node attention in addition to edge attention and using multi-headed attention or other attention pooling techniques. In our work, we manually analyzed self-attention visualizations, but as the field of neural network interpretation advances,\cite{GNN} future work on developing tools that improve and automate this process would make deep learning analysis of amorphous material structures even more effective. 

\begin{acknowledgments}
\noindent
The authors are grateful to Ashley Guo and Cody Bezik for useful discussions and comments and to Juan J. de Pablo for support. Shubhendu Trivedi's work was supported by the National Science Foundation under Grant No. DMS-1439786 while the author was in residence at the Institute for Computational and Experimental Research in Mathematics in Providence, RI, during the non-linear algebra and computer vision programs. Risi Kondor was partially supported by DARPA HR00111890038 and this project used computational resources provided by NSF MRI 1828629.
\end{acknowledgments}

\bibliography{manuscript}

\appendix 
\section{Simulation Details}
\par In each of the 20,000 simulations that we performed, the same initial configuration of particles is used as a starting point. This configuration is minimized using the FIRE minimization algorithm, which brings the configuration to its local potential energy minimum. \cite{F} To do this we used an energy stopping tolerance of $1\times 10^{-10}$, a force stopping tolerance of $1\times 10^{-5}$, a maximum of 1,000 iterations of the minimizer, and a maximum of 1,000 force and energy evaluations. To ensure independence of the simulations, the particles in this configuration are then given initial velocities drawn from a Gaussian distribution with mean zero and with a total linear momentum of zero and a temperature of 2.0. The seeds used to initialize random particle velocities in each of the 10,000 simulations are generated using the Bash \$RANDOM function, which returns a pseudorandom integer in the range 0 to 32,767.\footnote{See http://tldp.org/LDP/abs/html/randomvar.html.} 
\par Each simulation is then run at constant particle number, volume, and temperature (canonical NVT ensemble) with an integration timestep of $\Delta t = 0.005$ in Lennard-Jones units. For the duration of the simulation, we cool the system linearly from the initial temperature of 2.0 to a final temperature of 0.05 using a damping parameter of 0.5. The system is constrained to two dimensions during the simulation using the \textit{enforce2D} command in LAMMPS, and the linear momentum is zeroed in each dimension at every timestep. Particle configurations (both scaled and unscaled coordinates) are recorded every 100,000 steps. \footnote{Unscaled coordinates refer to the raw coordinates output by a simulation. Scaled coordinates refer to the raw coordinates normalized so that they lie within the unit interval [0, 1].}
\par After the simulation steps are complete, we measure the system's inherent structure energy, also at intervals of 100,000 steps, by loading in the scaled configurations output from the simulation.  To compute inherent structure energy, we again use the FIRE minimization algorithm as described above, this time with a maximum of 10,000,000 iterations and a maximum of 10,000,000 force and energy calculations.

\par
\begin{figure*}
\centering
\subfloat[]{
  \includegraphics[scale = 0.5]{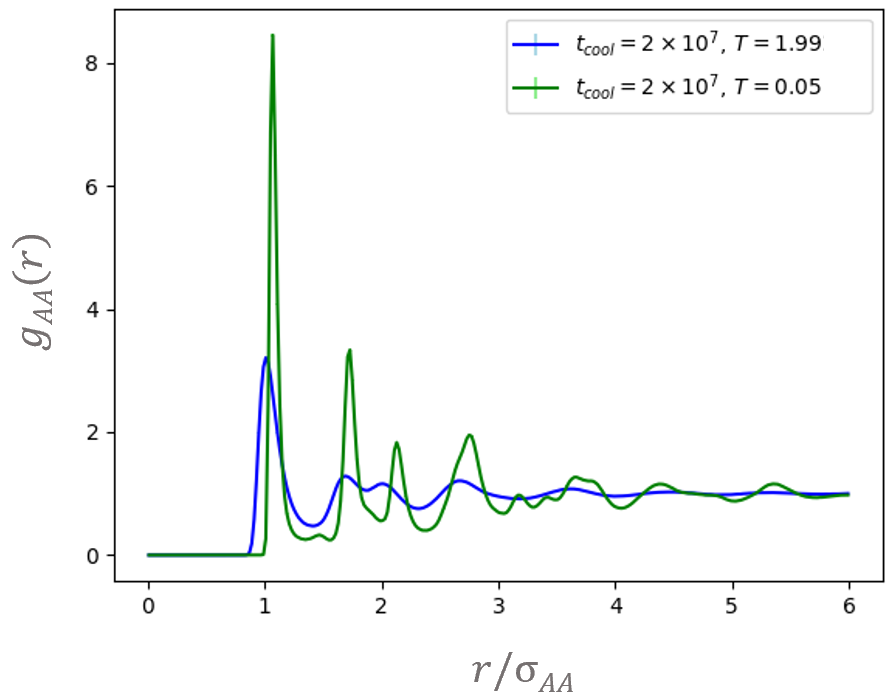}}
  
\subfloat[]{
  \includegraphics[scale = 0.5]{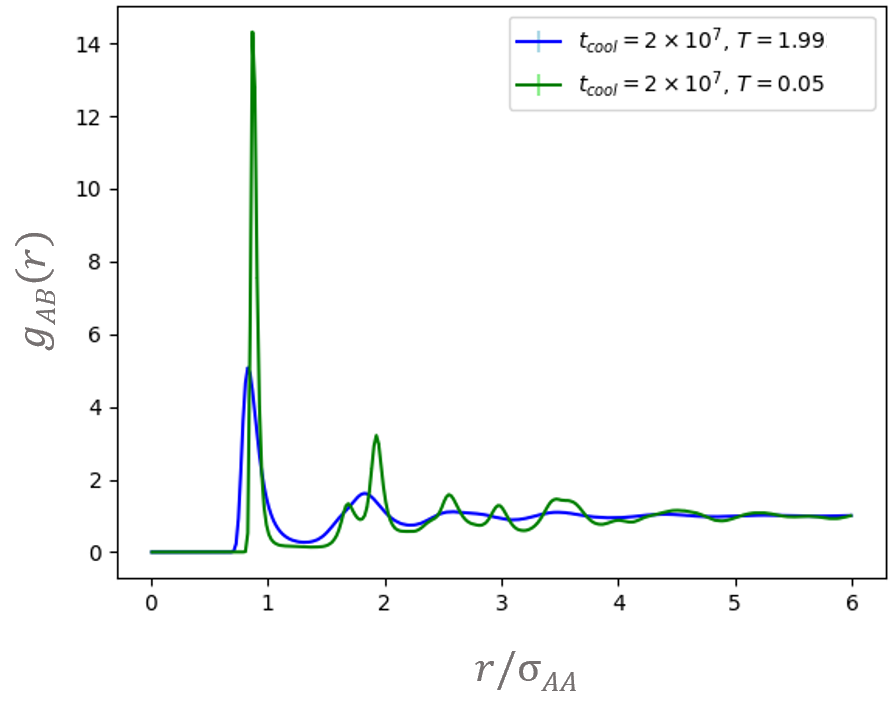}}
  
\subfloat[]{
  \includegraphics[scale = 0.5]{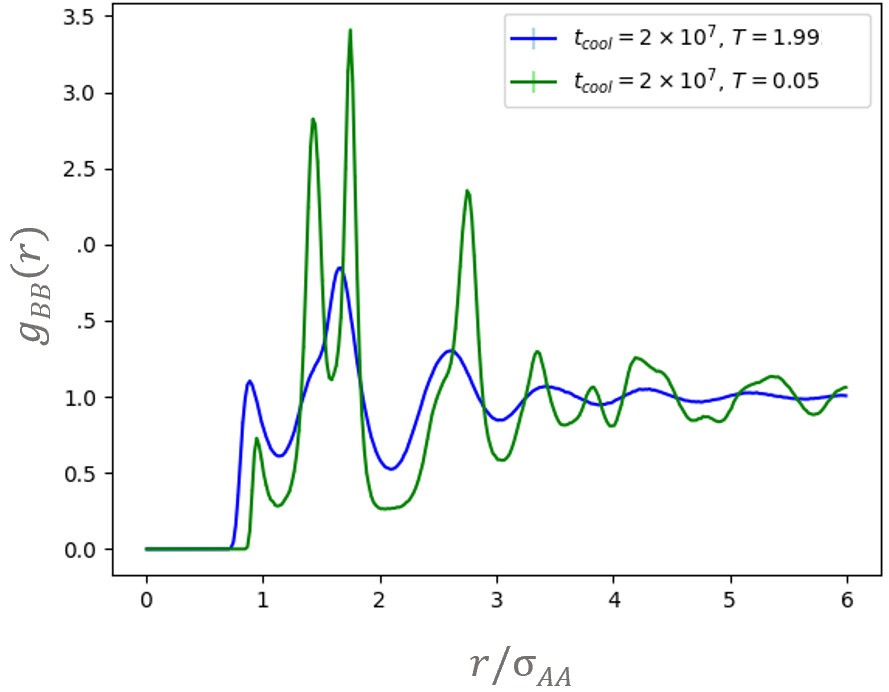}}
\caption{\label{fig:figure11} \small Average \textbf{(a)} A-A, \textbf{(b)} A-B, and \textbf{(c)} B-B radial distribution functions of 10,000 configurations at T = 1.99 and 10,000 configurations at T = 0.05 with $t_{cool}=2\times 10^7$ (dataset 1). Glasses (green) are clearly distinguishable from liquids (blue) because of their higher and sharper radial distribution function peaks.}
\end{figure*}

 Figure \ref{fig:figure11} shows comparisons of the average radial distribution functions of liquid and glass configurations in dataset 1, where the radial distribution function is defined as 

\begin{align}
\begin{split}
g_{ij}(r) = \frac{N}{\rho N_i N_j}\sum_{m = 1}^{N_i}\sum_{n=1}^{N_j}\left<\delta\left(\textbf{r} - \textbf{r}_{mn}\right)\right> \hspace{5mm} i, j\in\{A, B\}. 
\end{split}
\end{align}
Here, $N$ is the total number of particles, $N_i$ is the total number of type $i$ particles, $N_j$ is the total number of type $j$ particles, $\rho$ is the total density, $\delta$ is the delta functional, $\textbf{r}_{mn}$ is the vector from particle $m$ to particle $n$, and the average is taken over all vectors $\textbf{r}$ with magnitude $r$. We calculated the radial distribution functions in LAMMPS using 500 histogram bins and a cutoff distance of $10\sigma_{AA}$. As expected, these radial distribution functions exhibit clear structural differences between the liquid and glass configurations, with the glasses having higher and sharper peaks than the liquids.

At $T = 0.55$ the 10,000 configurations from $t_{cool}=2\times 10^7$, which are liquids, have a mean inherent structure energy of -3.86698 with a standard deviation of 0.00382, while at $T = 0.05$ the 10,000 configurations from $t_{cool} = 2\times 10^5$, which are glasses, have a mean inherent structure energy of -3.86744 with a standard deviation of 0.00288. These sets of configurations, therefore, have approximately the same energy.
\par
\begin{figure*}
\centering
\subfloat[]{
  \includegraphics[scale = 0.5]{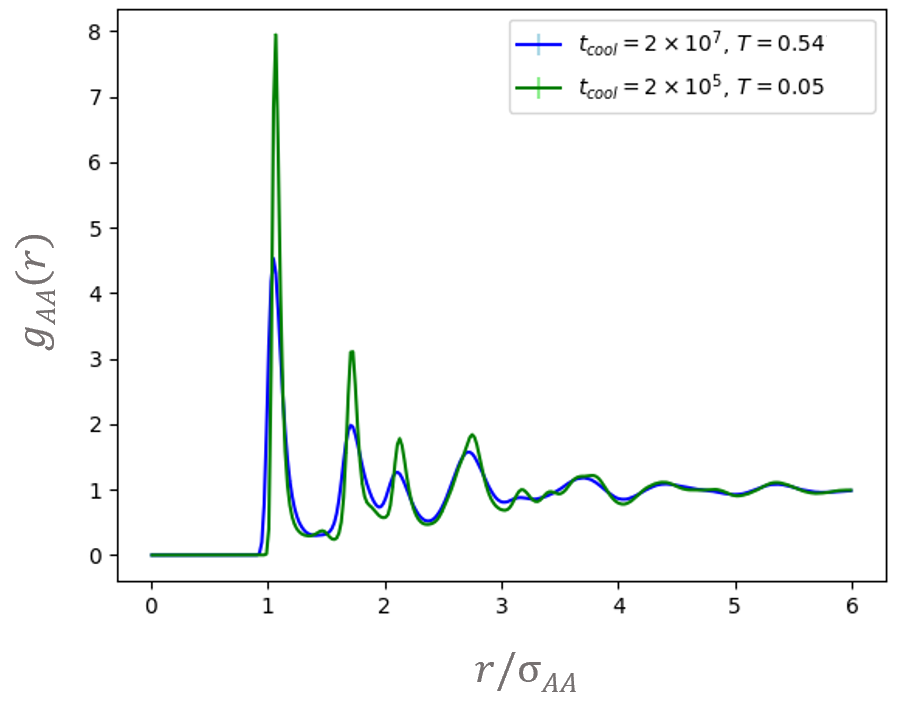}}
  
\subfloat[]{
  \includegraphics[scale = 0.5]{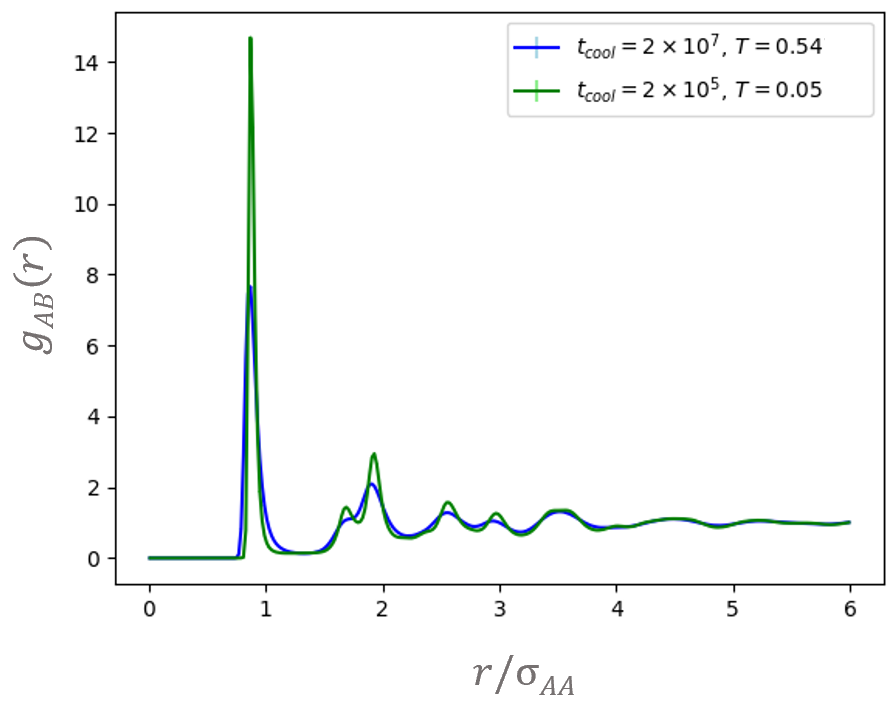}}
  
\subfloat[]{
  \includegraphics[scale = 0.5]{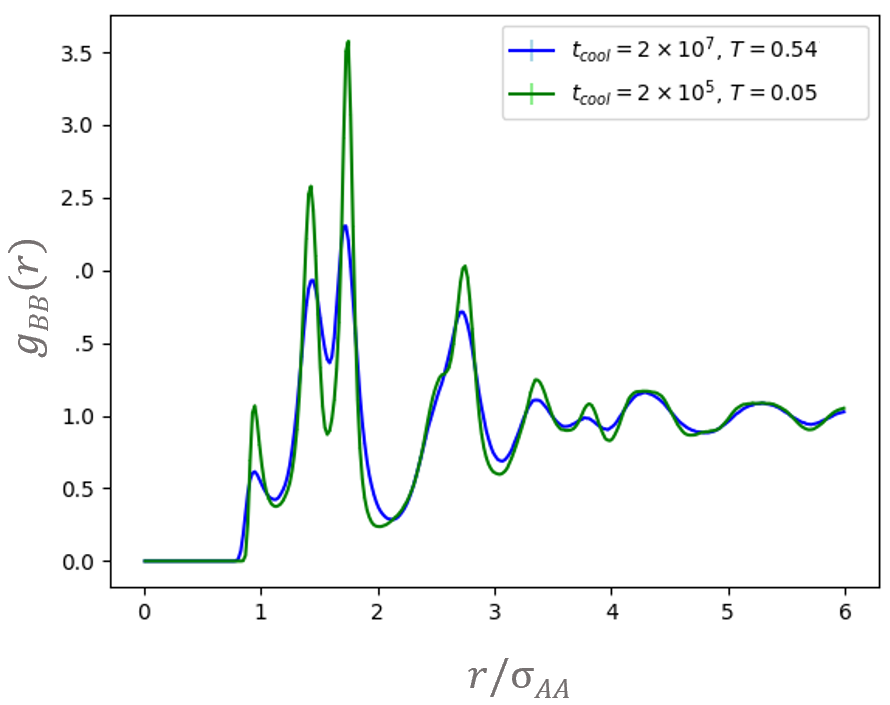}}
\caption{\label{fig:figure12} \small Average \textbf{(a)} A-A, \textbf{(b)} A-B, and \textbf{(C)} B-B radial distribution functions of 10,000 configurations at T = 0.55 with $t_{cool} = 2\times 10^7$ and 10,000 configurations at T = 0.05 with $t_{cool}=2\times 10^5$ (dataset 6). Glasses (green) are distinguishable from liquids (blue) because of higher and sharper radial distribution function peaks, but the differences are much less severe than in Figure \ref{fig:figure11} as the glasses and liquids here have the same average inherent structure energy.}
\end{figure*}

 We computed average radial distribution functions comparing the configurations from these glasses and liquids, which comprise dataset 6, shown in Figure \ref{fig:figure12}. Note that the structures are much more similar than those shown in Figure \ref{fig:figure11}.

\section{Image Data for CNNs} We utilized CNNs to classify liquid and glass configurations by rendering them as images. For a given configuration, we load its scaled particle coordinates and particle types as a \textit{numpy} array in Python and used the \textit{matplotlib} package to save the configuration as a 250 x 250 pixel PNG image with 100 dots per inch. As noted at https://lammps.sandia.gov/doc/dump.html: \textit{"Because periodic boundary conditions are enforced only on timesteps when neighbor lists are rebuilt, the coordinates of an atom written to a dump file may be slightly outside the simulation box...atom coords are written in a scaled format (from 0 to 1)...an x value of 0.25 means the atom is at a location 1/4 of the distance from xlo to xhi of the box boundaries."} Upon inspection, most particles have coordinate values between 0 and 1, with only occasional exceptions slightly outside these bounds, e.g. values such as 1.0001. This does not affect our analyses. We remove the axes and the frame from the image so that only the particles are rendered in the image, without any additional artifacts. Type A particles are represented as orange dots and type B particles as blue dots using the \textit{scatter} function, with a dot size of $s = 1$. The image is then slightly cropped in order to remove any unnecessary white space. In some images this slightly truncates particles at the edges of the image, but we found that this does not inhibit the performance of the CNN. Metadata representing these images for a dataset are saved in a JSON file.  
\par During training the generator function loads a batch of training images into a \textit{numpy} array by reading their file paths from the metadata JSON file. This function converts the liquid and glass labels into a one-hot vector representation, i.e. ([0, 1] for glasses and [1, 0] for liquids). 

\section{Hyperparameter Optimization for D-MPNNs}

\par The \textit{hyperopt} Sequential Model-based Global Optimization (SMBO) algorithms form a probabilistic model that maps hyperparameters to a probability of a score on the loss function, $P(y|x)$. This probabilistic model is called a surrogate.  SMBO methods work by choosing the next set of hyperparameters to test on the loss function by selecting hyperparameters that perform best on the surrogate function. \cite{B} As each set of hyperparameters is evaluated, the method updates the surrogate probability model in order to make increasingly well-informed guesses. After a specified number of iterations, the method suggests the optimal set of hyperparameters. The specific SMBO method that we use in this work is called a Tree-structured Parzen Estimator (TPE), which is thoroughly described in Bergstra \textit{et al.} and which we follow closely here. \cite{B}  There are different ways of identifying which hyperparameters to select based on the surrogate model in an SMBO method, but one of the most effective is a metric called Expected Improvement, otherwise known as an "exploration-exploitation" criterion.  Given a desired threshold value for the objective function, $y^*$, and some set of hyperparameters $x$, the Expected Improvement is given by 

\begin{align}
\begin{split}
\text{EI}_{y^*}(x) = \int_{-\infty}^{y^*}(y^* - y)P(y|x)dy.
\end{split}
\end{align}
The first factor in the integrand promotes values in regions that are likely to contain objective function minima (exploitation), while the second term promotes regions that have greater uncertainty (exploration).  When this integral is positive, it means that the hyperparameter set $x$ is expected to yield an improvement relative to the threshold value $y^*$.
\par In the TPE algorithm, instead of modeling the surrogate directly as $p(y|x)$, this method uses Bayes rule, $p(y|x) = \frac{p(x|y)p(y)}{p(x)}$, to model $p(x|y)$ and $p(y)$ instead.  $p(x|y)$ is broken down into $l(x)$ and $g(x)$, such that 

\begin{align} p(x|y) \ := \   \begin{cases}
       l(x)  & y < y^* \\
       g(x) & y \ge y^*.
     \end{cases}
\end{align}   
In other words, we create two different distributions for the hyperparameters: one where the objective function value is less than the threshold, $l(x)$, and one where the objective function value is greater than the threshold, $g(x)$.  These non-parametric densities are constructed after some number $K$ of evaluations of the objective function.  $y^*$ is chosen to be slightly greater than the best observed objective function score.    

In this approach, the Expected Improvement is given by 

\begin{align}
\begin{split}
\text{EI}_{y^*}(x) = \int_{-\infty}^{y^*}(y^* - y)\frac{p(x|y)p(y)}{p(x)}dy,
\end{split}
\end{align} 
which can be rearranged as 

\begin{align}
\begin{split}
\text{EI}_{y^*}(x) \propto \left(\gamma + \frac{g(x)}{l(x)}(1 - \gamma)\right)^{-1},
\end{split}
\end{align}
where $\gamma = p(y < y^*)$ (no specific $p(y)$ is necessary).  So, the TPE works by drawing sample hyperparameters from $l(x)$, evaluating them in terms of $g(x)/l(x)$, and returning the set $x$ that gives the best expected improvement value.

\section{Attempts to Interpret CNNs} 
\begin{figure*}
\subfloat[]{
  \includegraphics[scale = 1.0]{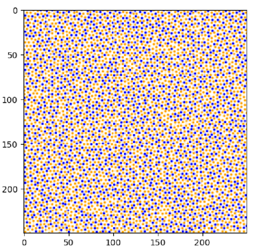}}%
\subfloat[]{
  \includegraphics[scale = 1.0]{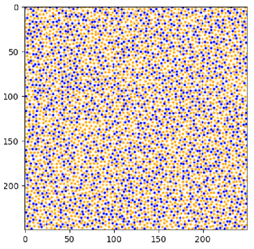}}%
\subfloat[]{
  \includegraphics[scale = 1.0]{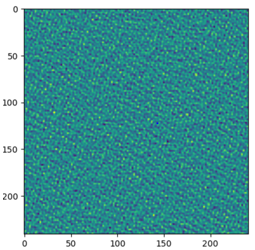}}
\subfloat[]{
  \includegraphics[scale = 1.0]{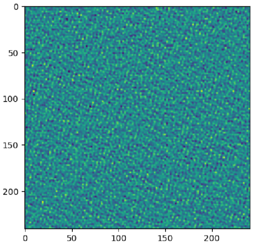}}%
  
\subfloat[]{
  \includegraphics[scale = 1.0]{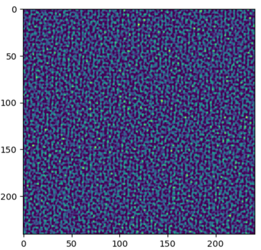}}%
\subfloat[]{
  \includegraphics[scale = 1.0]{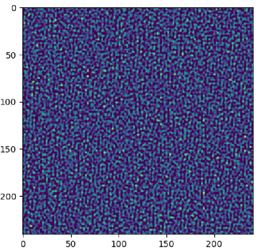}}%
\subfloat[]{
  \includegraphics[scale = 1.0]{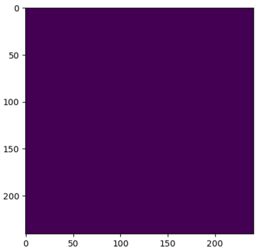}}
\subfloat[]{
  \includegraphics[scale = 1.0]{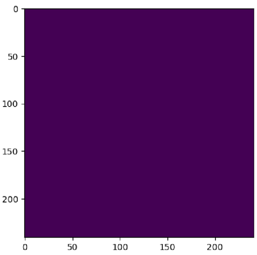}}%
  
\subfloat[]{
  \includegraphics[scale = 1.0]{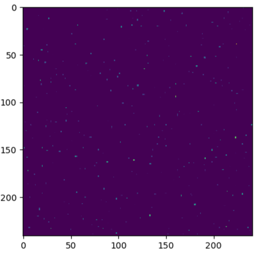}}%
\subfloat[]{
  \includegraphics[scale = 1.0]{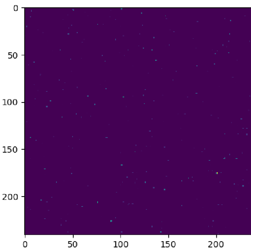}}%
\subfloat[]{
  \includegraphics[scale = 1.0]{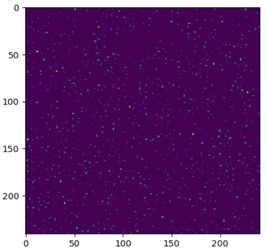}}
\subfloat[]{
  \includegraphics[scale = 1.0]{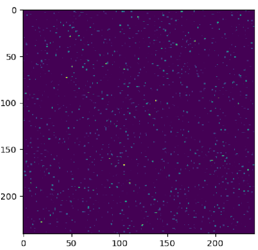}}%
  
\subfloat[]{
  \includegraphics[scale = 1.0]{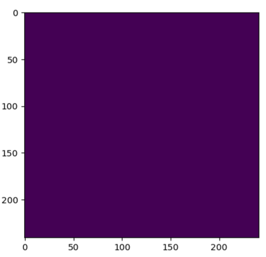}}
\subfloat[]{
  \includegraphics[scale = 1.0]{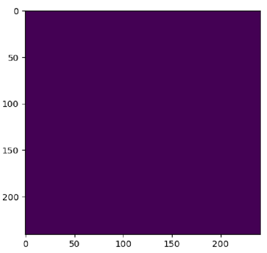}}%
\caption{\label{fig:figure13}\small First convolutional layer feature map visualizations for representative liquid and glass configurations. The network used here is a model optimized on dataset 1. The glass and liquid images are taken from $t_{cool}= 2\times 10^7$ data and were correctly classified by the network. The axes on all of these plots correspond to pixel number. \textbf{(a)} is the original glass configuration; \textbf{(b)} is the original liquid configuration; \textbf{(c)} is glass feature map 1; \textbf{(d)} is liquid feature map 1; \textbf{(e)} is glass feature map 2; \textbf{(f)} is liquid feature map 2; \textbf{(f)} is glass feature map 3;
\textbf{(h)} is liquid feature map 3; \textbf{(f)} is glass feature map 4;
\textbf{(h)} is liquid feature map 4; \textbf{(f)} is glass feature map 5;
\textbf{(h)} is liquid feature map 5; \textbf{(f)} is glass feature map 6;
\textbf{(h)} is liquid feature map 6.}
\end{figure*}

In an attempt to interpret the CNNs, we computed the feature maps produced by the first convolutional layer for several glass and liquid images that were correctly classified as such by the best network trained on dataset 1. Visualizing components of neural networks is gaining traction as a method for interpretation \footnote{For example, see https://distill.pub/2017/feature-visualization/.}. Visualizations of these feature maps for representative liquid and glass configurations are shown in Figure \ref{fig:figure13}.  \\

These visualizations are not clearly interpretable. Feature maps 3 (Figures \ref{fig:figure13}(g) and \ref{fig:figure13}(h)) and 6 (Figures \ref{fig:figure13}(m) and \ref{fig:figure13}(n)) are blank. Feature maps 1 (Figures \ref{fig:figure13}(c) and \ref{fig:figure13}(d)) and 2 (Figures \ref{fig:figure13}(e) and \ref{fig:figure13}(f)) appear to show some kind of texture pattern, with map 1 having bumps that are visually reminiscent of regions of the original images concentrated with type A (orange) particles. Feature maps 4 (Figures \ref{fig:figure13}(i) and \ref{fig:figure13}(j)) and 5 (Figures \ref{fig:figure13}(k) and \ref{fig:figure13}(l)) appear to highlight specific pixels in the image. Perhaps these correspond to specific particles from the original images, but they do not map back directly to pixels that correspond to specific particle locations. 
\par We attempted a different method for interpretation based on our prior knowledge about local geometric structure in two-dimensional Kobb-Anderson binary mixtures. As evidenced in Reid \textit{et al.}, the degree of five-fold symmetry in a liquid-cooled glass is much higher than that in liquids at higher temperatures and higher inherent structure energies. \cite{R} One possibility is that the network has identified this geometric quantity as a means for classifying liquids and glasses and that the kernels in the CNN have been trained to identify local pentagonal arrangements of particles. As in Reid \textit{et al.}, we consider a region of five-fold symmetry to be characterized by a type B (blue) particle surrounded immediately by a pentagon of type A (orange) particles. Perhaps, if the CNN is trained to correlate high concentrations of five-fold symmetry with glassy materials, an artificial image that is saturated with five-fold symmetry patterns will be classified by the network as a glass.  
\par
\begin{figure}
\subfloat[]{
  \includegraphics[scale = 0.55]{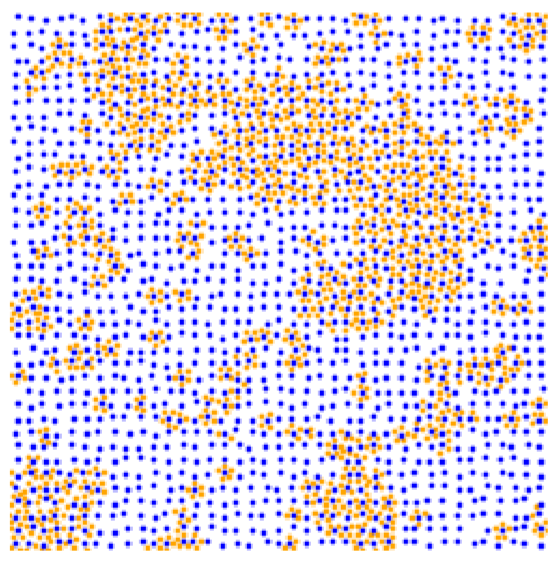}}%
\subfloat[]{
  \includegraphics[scale = 0.55]{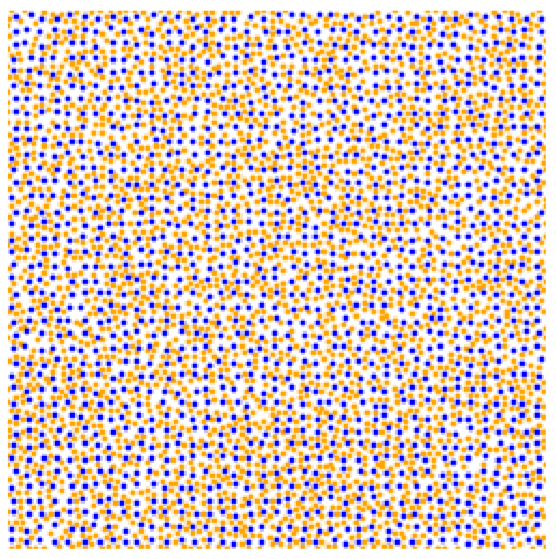}}%
\caption{\label{fig:figure14}\small \textbf{(a)} Initial artificial pentagonal arrangements of type A and type B particles. \textbf{(b)} Final artificial five-fold symmetry configuration.}
\end{figure}

 We tested this hypothesis by constructing artificial images imbued with five-fold symmetry as follows. First, we placed 1,512 (35\% of 4,320) type B particles in a grid, evenly spaced, with the x- and y-axes ranging from 0 to 1. Each particle was then given a slight random displacement in the x- and y-directions from their initial placements, corresponding to a uniform random number in the range -1/195 to 1/195.\footnote{These specific random displacement values were determined by visual inspection and trial and error.} We then selected 500 of these type B particles to surround with pentagonal arrangements of type A (orange) particles. To attempt to replicate the five-fold symmetry clustering effect described in Hu \textit{et al.}, we added pentagonal arrangements around type B particles iteratively and selected each subsequent type B particle to be a neighbor of a previously selected particle with a probability of 3/4.\cite{N} Every time a pentagonal arrangement of type A particles was placed, we arranged the particles to be at a radial distance of 1/78 from the central type B particle and gave the pentagonal arrangement a random angular rotation selected uniformly from the range $-2\pi \text{ to } 2\pi$. For any pair of particles within a distance of 1/90 of each other, we removed one of the particles to prevent overlaps. Figure \ref{fig:figure14}(a) shows a representative image of the result.

The remainder of the type A particles are filled in randomly, again avoiding any overlaps, to produce a final result, shown in \ref{fig:figure14}(b). We again used the dataset 1 network to classify several hundred examples of these artificial images. However, all of them were classified as liquids. We also constructed images with other n-fold symmetries and with random configurations of particles, but all were classified as liquids.  

\section{CNN and D-MPNN Classification Results}
\begin{figure}
\includegraphics[scale = 0.8]{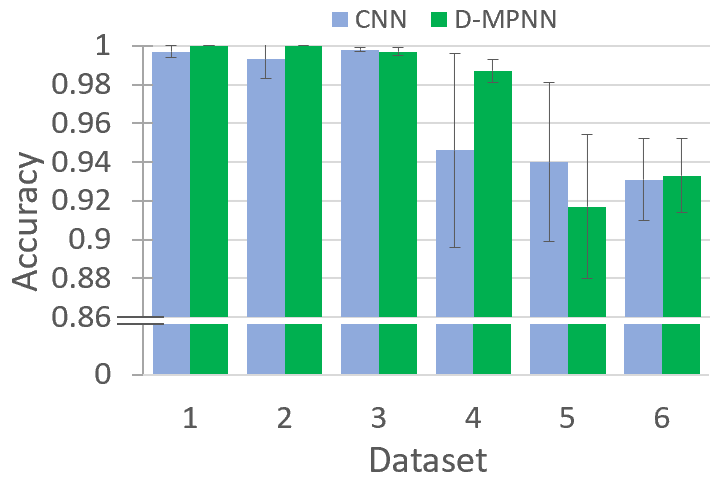}
\caption{\label{fig:figure15} \small CNN and D-MPNN average classification accuracy for datasets 1 through 6 with error bars showing standard deviation. These average values were computed using the three-fold nested cross-validation scheme described in \S\ref{sec:sec25}.}
\end{figure}

Figure \ref{fig:figure15} shows accuracy results for CNNs and D-MPNNs.

\section{D-MPNN Self-Attention}
\begin{figure}
\centering
\subfloat[]{
  \includegraphics[scale = 0.49]{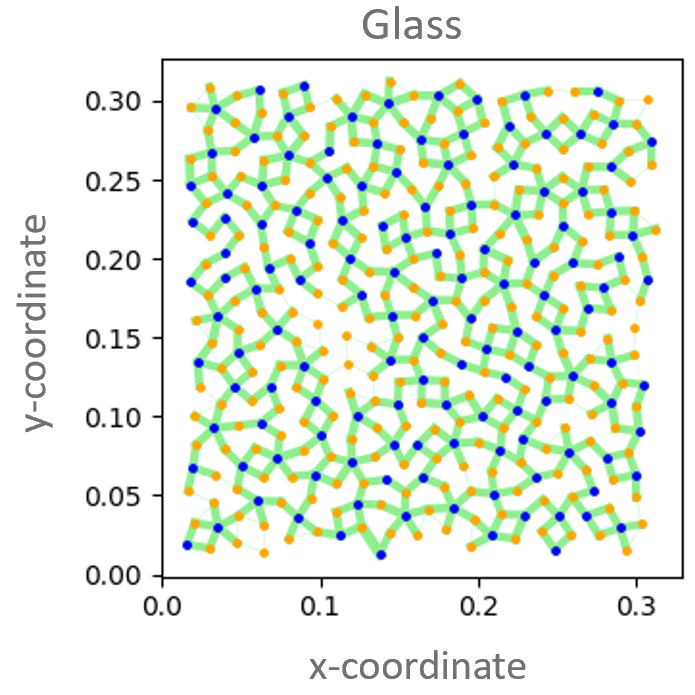}}%
\subfloat[]{
  \includegraphics[scale = 0.49]{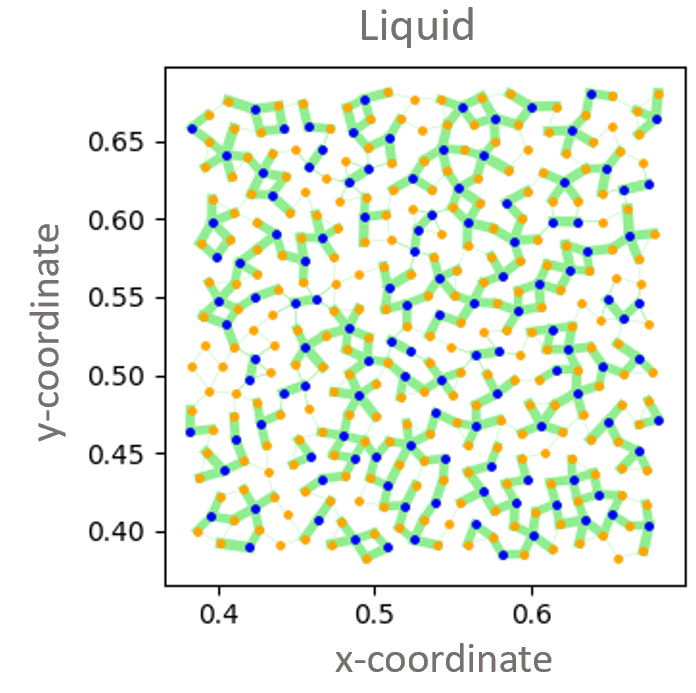}}%
\caption{\label{fig:figure16} \small Self-attention visualizations, with the attention weights computed on the $\mathcal{T} - 1^{th}$ step of message passing. All connected particles in the graph are joined with a green line whose width is proportional to the magnitude of the attention weight. Note that each connected pair of particles actually has two edges, because the graph is directed. Here, we visualize the edge with the higher weight. \textbf{(a)} Glass configuration with $l_{win} = 0.3$. \textbf{(b)} Liquid configuration with $l_{win} = 0.3$.}
\end{figure}

\begin{figure*}
\centering
\subfloat[]{
  \includegraphics[scale = 0.5]{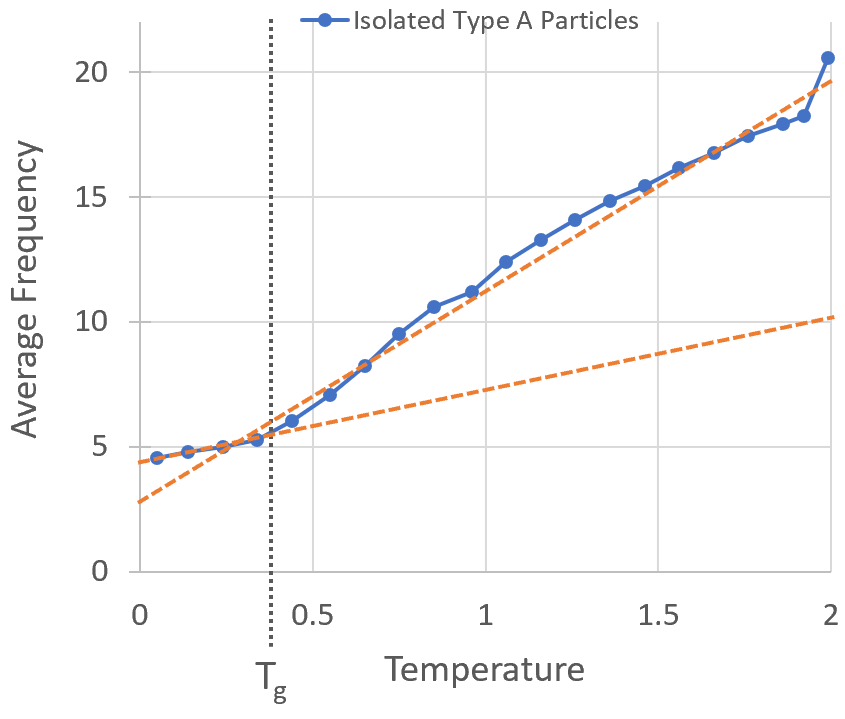}}
  
\subfloat[]{
  \includegraphics[scale = 0.5]{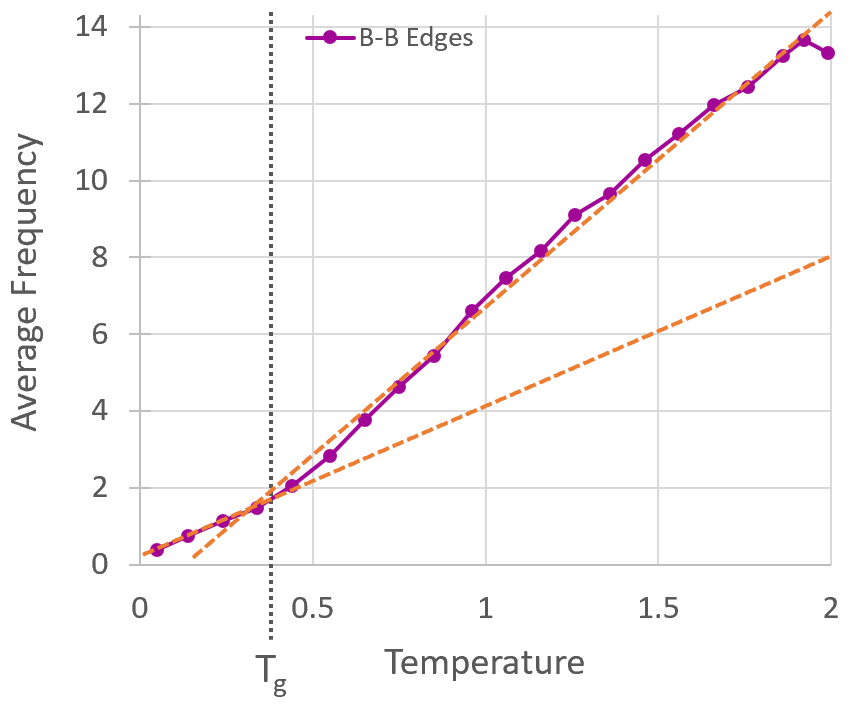}}
  
\subfloat[]{
  \includegraphics[scale = 0.5]{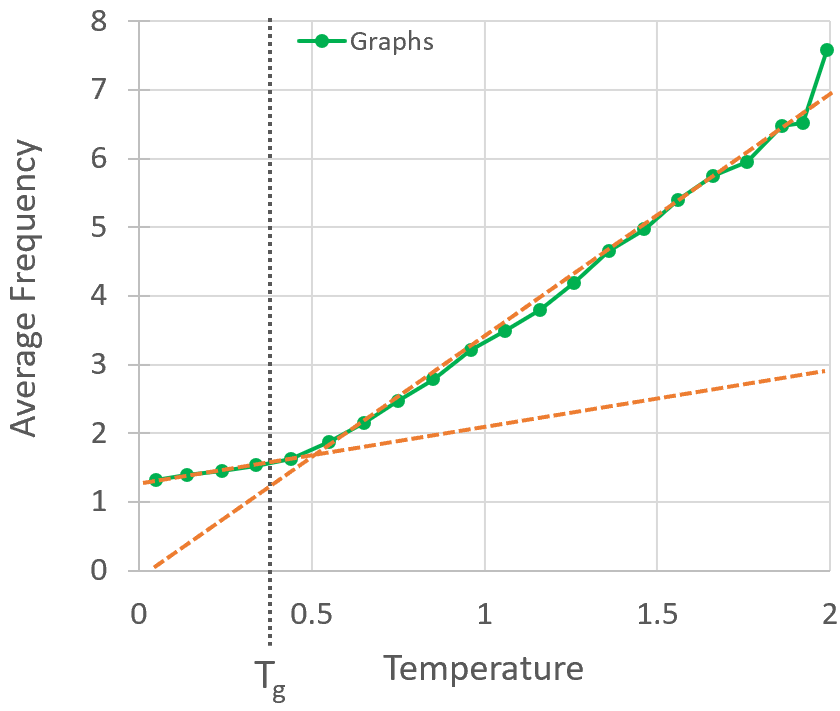}}
\caption{\label{fig:figure17} \small Average number of isolated type A particles \textbf{(a)}, edges connecting pairs of type B particles, \textbf{(b)}, and number of disjoint graphs \textbf{(c)} in configurations from simulations with $t_{cool}=2\times 10^7$ at a variety of temperatures. The dotted orange lines show predictions from the linear regression models for each of these curves, as described in \S\ref{sec:sec3}. The segments of the linear models above and below $T_g$ are artificially extended to visually highlight the difference in slopes above and below $T_g$.}
\end{figure*}

Figure \ref{fig:figure16} shows self-attention visualizations for configurations in an outer test set of dataset 1 with $l_{win}=0.3$. High attention edges were determined with a hard cutoff; upon inspection, there was a clear trough in the distribution of attention weights that separated those with small, almost negligible magnitudes and those with larger magnitudes. Figure \ref{fig:figure17} shows the graph-based metrics from Figure \ref{fig:figure10} plotted individually as a function of temperature.

\end{document}